\documentclass[letterpaper, 10 pt, conference]{ieeeconf}  

\IEEEoverridecommandlockouts                              

\usepackage{graphics} 
\usepackage{epsfig} 
\usepackage{amsmath} 
\usepackage{amssymb}  
\usepackage{amsfonts}  
\usepackage{todonotes}

\usepackage{xspace}
\usepackage{soul}
\usepackage{mathtools}
\usepackage{makecell}
\usepackage{nicefrac}
\usepackage{float}
\usepackage{wrapfig}

\newcommand{\ra}{\rightarrow}

\newcommand{\R}{\ensuremath{\mathbb{R}}}

\newcommand{\X}{\ensuremath{\mathbb{X}}}

\newcommand{\U}{\ensuremath{\mathbb{U}}}

\newcommand{\sm}{\ensuremath{\setminus}}

\let\emptyset\varnothing

\newcommand{\eps}{\ensuremath{\varepsilon}}

\newcommand{\ball}{\ensuremath{\mathcal B}}

\newcommand{\abs}[1]{\left\lvert#1\right\rvert}
\newcommand{\scal}[1]{\left\langle#1\right\rangle}

\definecolor{dgreen}{rgb}{0.0, 0.5, 0.0}

\usepackage{textcomp}

\makeatletter
\newcommand{\subalign}[1]{%
	\vcenter{%
		\Let@ \restore@math@cr \default@tag
		\baselineskip\fontdimen10 \scriptfont\tw@
		\advance\baselineskip\fontdimen12 \scriptfont\tw@
		\lineskip\thr@@\fontdimen8 \scriptfont\thr@@
		\lineskiplimit\lineskip
		\ialign{\hfil$\m@th\scriptstyle##$&$\m@th\scriptstyle{}##$\crcr
			#1\crcr
		}%
	}
}

\usepackage{xcolor}
\usepackage{subcaption}
\usepackage{siunitx}

\newtheorem{thm}{Theorem}

\newtheorem{lem}{Lemma}

\newtheorem{asm}{Assumption}
\newtheorem{rem}{Remark}

\newcommand{\norm}[1]{\left\lVert#1\right\rVert}  
\newcommand{\set}[1]{\mathbb #1}

\usepackage{algorithm, algorithmic}

\title{\LARGE \bf Some remarks on practical stabilization via CLF-based control under measurement noise}

\author{Patrick Schmidt, Pavel Osinenko, Stefan Streif
\thanks{Patrick Schmidt and Stefan Streif are with Technische Universit\"at Chemnitz, 09126 Chemnitz, Germany, Automatic Control and System Dynamics Lab (e-mail: \{patrick.schmidt, stefan.streif\}@etit.tu-chemnitz.de).}
\thanks{Pavel Osinenko is with Skolkovo Institute of Science and Technology, 143026 Moscow, Russia (e-mail: p.osinenko@yandex.ru)}
\thanks{This work has been accepted in the IEEE Access, 10.1109/ACCESS.2024.3521048.}
\thanks{©\the\year IEEE. Personal use of this material is permitted. Permission from IEEE must be obtained for all other uses, in any current or future media, including reprinting/republishing this material for advertising or promotional purposes, creating new collective works, for resale or redistribution to servers or lists, or reuse of any copyrighted component of this work in other works.
}
}
\begin{document}
\maketitle
\thispagestyle{empty}
\pagestyle{empty}

\begin{abstract}                
	Practical stabilization of input-affine systems in the presence of measurement errors and input constraints is considered in this brief note.
Assuming that a Lyapunov function and a stabilizing control exist for an input-affine system, the required measurement accuracy at each point of the state space is computed.
This is done via the Lyapunov function-based decay condition, which describes along with the input constraints a set of admissible controls.
Afterwards, the measurement time points are computed based on the system dynamics.
It is shown that between these self-triggered measurement time points, the system evolves and converges into the so-called target ball, i.e. a vicinity of the origin, where it remains.
Furthermore, it is shown that the approach ensures the existence of a control law, which is admissible for all possible states and it introduces a connection between measurement time points, measurement accuracy, target ball, and decay.
The results of the approach are shown in three examples.
\end{abstract}

\begin{keywords}
	Control Lyapunov function, measurement noise, practical stabilization, self-triggered control
\end{keywords}

\section{Introduction} \label{sec:intro}

Measuring always comes along with measurement errors.
This causes a discrepancy between the real output of the system and the measurement, which propagates to the estimated states.
This discrepancy of the states is simply denoted as measurement error from now on, since the estimation of states is not considered here.
However, the framework can be also considered for errors in state estimation.

Measurement errors might result in diverging closed-loop trajectories.
If the measurement error is \emph{small enough}, \cite{Sontag1999-stabilization-disturb} presented results that the system behavior is similar if a continuously differentiable feedback is applied to the system, which is evaluated on the measured state.
However, a bound on the measurement error is missing in most publications.
The aim of the current publication lies in contributing approaches to close this research gap by investigating different terms that influence the measurement error and to derive bounds for them.

In general, dealing with imprecise state measurements is a hard problem.
Several works consider specific examples, e.g., \cite{Mazenc2019-stabilization} and references therein, where imprecise measurements for landing an aircraft are considered.
More general results are often conservative.
They exist both for linear \cite{Behera2014-event} and nonlinear systems \cite{Liberzon2005-stabilization}.
Stabilization of nonlinear systems via nonsmooth Lyapunov functions under measurement errors is considered in \cite{Ledyaev1997-remark} and \cite{Sontag1999-stabilization-disturb}.
In \cite{Schmidt2021-inf}, stabilization of nonlinear systems is investigated under the presence of measurement noise and system disturbance. 
The authors provide bounds for the sampling time of a sample-and-hold control and for the disturbances.
In this note, the latter approach is considered as a basis, since it considers practical stabilization in the presence of measurement errors.

The idea presented here consists of stabilizing all possible states within a ball around the actual state, i.e. all states that can be obtained by a measurement, where the radius of this ball describes the given measurement error.
The issue is that the magnitude of the measurement error defines whether stabilization is possible or not, since the larger this magnitude becomes, the more states are possible that all have to be stabilized by the same control.
If this magnitude is too large, there might be no admissible control, where the magnitude as well as the set of admissible controls depend on the current state.
Choosing a state-independent measurement error bound results in a conservative choice of this magnitude.
However, at some states there is a time period in which the system evolves until the set of admissible controls is empty.
Before this set becomes empty, a new measurement has to be made, since the discrepancy between current state and measured states is too large.
For this task, a triggering rule is required.

In literature, there are different triggering rules.
A quite conservative, but wide-spread approach is a time-triggered measurement, where a new measurement is obtained after a specified time.
In the last years, a lot of effort was put especially into event-triggered control (see \cite{Borgers2018-time}, \cite{lin2024artificial}, \cite{Scheres2024-robustifying}, \cite{Tarbouriech2018-insights} and references therein) as well as self-triggered control.
The latter approach is considered for linear systems (\cite{Brunner2019-event}, \cite{Heemels2012-introduction}, \cite{Liu2023-data}, \cite{Rajan2023-analysis}, \cite{Zhang2022-finite}), but also for nonlinear systems (\cite{Anta2010-sample}, \cite{Gommans2015-resource}, \cite{Proskurnikov2019-lyapunov}) even with delays (\cite{Hertneck2021-dynamic}, \cite{Hertneck2022-dynamic}), safety requirements \cite{Kishida2023-greedy}, different perturbations, as well as parameter uncertainties \cite{DiBenedetto2013-digital}.
In event-triggered control (ETC), a triggering condition is monitored.
Once it is violated, the event is triggered.
In contrast, if the next update time is computed based on predictions using available state information and knowledge of the plant dynamics, it is called self-triggered control (STC) \cite{Hertneck2022-dynamic}.
Heemels et al. provide a survey on ETC and STC in \cite{Heemels2012-introduction} and \cite{Heemels2021-event}.
The main difference of ETC and STC is the point in time when a control is computed:
Event-triggered control is reactive, i.e. it generates a control once the states of the system deviate more than a certain threshold from a desired value.
However, self-triggered control is proactive and computes the next control ahead of time \cite{Heemels2021-event}.
The paper focuses on STC due to its proactive character, where a function $\delta_k: \R^n \ra \R_+$ is determined that computes the time point to obtain the next measurement based on the actual state.
STC is applied in a vast field of applications, e.g. autonomous driving \cite{Sabouni2024-optimal}, multi-agent systems \cite{Li2019-event, Yi2018-dynamic}, mobile robots \cite{Zhang2022-tracking}, and networked control systems \cite{Hertneck2023-self} just to name a few.

However, STC might result in frequent measurements and a conservative control law, if an admissible control is computed and kept constant until it is not admissible anymore, which means that it does not stabilize all the states in the considered surrounding.
Therefore, the control is not kept constant between two triggering time points, which ensures that it is admissible after a new measurement and that no points of discontinuity of the control input appear, which results in a closer look on the system if such a control strategy is applied (cf. \cite{Li2001-switched}).

This brief note includes two contributions.
First, a condition for the sensor measurement accuracy is derived at each point in the state space, such that there exists an input that ensures a decay for all points around the current measurement.
This state-dependent measurement accuracy yields a link between the given measurement error and the target ball, i.e. the ball into which the closed-loop trajectories converge.
Second, a triggering rule is presented such that the closed-loop input-affine system converges into this target ball around the origin and remains there.
With these investigations on the interplay between different parameters, it allows a quantification of these parameters and it answers the initial question how precise the measurement has to be in order to guarantee stability.
The procedure of determining the triggering rules follows a different procedure than other existing approaches, which underlines the novelty of the presented approach.
For example in \cite{Anta2010-sample}, the triggering rule is based on state-dependent homogeneity.
In \cite{Proskurnikov2019-lyapunov}, the existence of a stabilizing self-triggered controller with a known convergence rate is shown.
In \cite{Hertneck2024-robust}, the triggering condition is chosen based on an upper bound of the evolution of the control Lyapunov function for nonlinear systems with disturbances (see also \cite{Mazo2009-self}).

Self-triggered control in the presence of measurement errors or uncertainties is, for example, considered in \cite{Delimpaltadakis2021-region}, where bounded disturbances and model uncertainties of the nonlinear system are tackled via differential inclusions that are included in known ETC/STC schemes.
In \cite{Zhang2021-new}, the authors consider drift uncertain nonlinearities along with nonholonomic constraints and they present a STC scheme based on integrating set-valued maps with state-separation and state-scaling techniques.
Linear systems along with an additive noise are considered in \cite{Kogel2015-robust} along with model predictive control (MPC).

The rest of the paper is organized as follows:
The problem setting is introduced in Section \ref{sec:prelim}.
Section \ref{sec:required-meas-acc} considers the set of admissible controls and visualizes it based on the choice of the measurement error.
Moreover, the maximum required measurement error is computed along with a lower bound, that depends on the target ball.
Section \ref{sec:triggering-condition} derives a triggering condition whenever a new measurement has to be taken.
Furthermore, it is shown in the main theorem that the closed-loop trajectories converge and remain in the target ball.
Finally, in Section \ref{sec:casestudy} the proposed approach is applied in a case study before the paper is concluded in Section \ref{sec:concl}.

Notation: $\ball_\eps(\hat x) := \{ x \in \R^n: \norm{x - \hat x} \leq \eps \}$ is defined as the ball of radius $\eps$ around $\hat x$, where $\norm{\cdot}$ is the Euclidean norm.
The Jacobian of a function $V: \R^n \ra \R$ is denoted as $\nabla_x V(x)$ and $\{ 1:m \} := \{ 1, 2, \ldots, m\}$.

\section{Problem description} \label{sec:prelim}

Consider an input-affine system
\begin{equation} \label{eqn:system}
	\dot x = f(x) + g(x)u, \quad x(0) = x_0
\end{equation}
with states $x \in \R^n$ and controls $u \in \R^m$.
The internal dynamics model $f: \R^n \ra \R^n$ and the input coupling function $g: \R^n \ra \R^{n \times m}$ are assumed to be continuously differentiable in $x$.
The input is bounded by box constraints, i.e., $\U := [u_{1,\min}, u_{1,\max}] \times [u_{2,\min}, u_{2,\max}] \times \ldots \times [u_{m,\min}, u_{m,\max}]$ with $0 \in \U$.
Those input constraints can be also understand as actuator limits.
To treat time-variant systems, the time can be added as an additional state, such that the resulting system reads as $\dot x = f(x; t) + g(x; t) u, \dot t = 1$.
For the ease of exposition, the paper focuses on time-invariant systems \eqref{eqn:system}.

Furthermore, a stabilizing controller $\kappa$ along with a CLF, which can be determined both numerically \cite{Baier2012-linear}, \cite{Baier2014-num-CLF}, \cite{Giesl2015-review} and analytically \cite{Bianchini2018-merging},
for \eqref{eqn:system} is assumed as follows
\begin{asm} \label{asm:decay-condition}
	There exists a continuously differentiable feedback controller $\kappa(x) \in \U$ and a twice continuously differentiable radially unbounded Lyapunov function $V$ such that for all $x \in \R^n$, it holds that 
	\begin{equation} \label{eqn:decay-condition-dot-V}
	\dot V(x) = \scal{\nabla_x V(x), f(x) + g(x) \kappa(x)} \leq - w(x), 
	\end{equation}
	where $w: \R^n \ra \R_{\geq 0}$ is continuously differentiable and satisfies $w(0) = 0$ and $x \not = 0 \implies w(x) > 0$.
\end{asm}

If $\kappa$ is to be applied, the current state must be known.
To determine the state, measurements are performed, where the measured state is denoted as $\hat x$.
The system is assumed to evolve based on the measurement $\hat x$ as follows:
\begin{equation} \label{eqn:system-hat}
	\dot{\hat x} = f(\hat x) + g(\hat x)u, \quad \hat x(0) = \hat x_0,
\end{equation}
where $f$ and $g$ were already introduced and $u = \kappa(\hat x)$ is used to obtain a decay of \eqref{eqn:system-hat} in the sense of Assumption \ref{asm:decay-condition}.
To obtain measurements, sensors are used, which, in practice, can only measure up to a certain precision.
This imprecise measurement results in deviations on the actual states such that the initial values of \eqref{eqn:system} and \eqref{eqn:system-hat} are assumed to fulfill $\norm{x_0 - \hat x_0} \leq \eps$, where $\eps$ is the accuracy of the measurement.
 
The idea is to stabilize \eqref{eqn:system} by stabilizing \eqref{eqn:system-hat} as well as all other states that could be obtained by the same measurement, i.e. all states in $\ball_\eps(x)$.
Since $\kappa$ is continuously differentiable, applying $\kappa(\hat x)$ to \eqref{eqn:system} results in a behavior which remains close to the intended one \cite{Sontag1999-stabilization-disturb}.
This property of robustness to small measurement errors is used to investigate the maximum admissible measurement error at each point $\bar \eps(x)$.
Based on this state-dependent admissible measurement error, minimum conditions on the sensor's accuracy $\eps$ are given such that the closed-loop trajectories converge into a vicinity of the origin.
This procedure is considered in Section \ref{sec:required-meas-acc}.
To prevent that the closed-loop trajectories diverge, measurements are performed whenever there exists no admissible control such that all possible states around the actual state are stabilized.
Such a condition that triggers a new measurement is determined in Section \ref{sec:triggering-condition}.

The procedure is summarized as follows:
First, the state $x_k := x(t_k)$ at $t = t_k$ is measures with an accuracy $\eps$.
This measurement yields $\hat x_k$.
Second, a feedback is determined, that stabilizes all points within the ball around the actual state based on the measurement $\hat x_k$, i.e. all $\tilde x \in \ball_{\eps}(x_k)$.
Third, the new measurement time point $t_{k + 1}$ is computed.
Once these steps are passed through, the procedure is repeated until the states of \eqref{eqn:system} are close to the origin, i.e. they are located in the so-called target ball.

The next section determines the required measurement accuracy $\bar \eps(x)$ at each point along with the minimum conditions on $\eps$.

\section{Computation of maximum measurement error} \label{sec:required-meas-acc}

In this section, a condition on the measurement accuracy is derived such that the decay condition \eqref{eqn:decay-condition-dot-V} holds for all points in a ball of radius $\eps$ around the current state.
First of all, a closer look at the decay condition is necessary.

\subsection{Visualization of constraints}

Based on \eqref{eqn:decay-condition-dot-V}, the decay condition for $u$ is defined as
\begin{equation} \label{eqn:decay-cond}
	\varphi(u,x; w) := \scal{\nabla_x V(x), f(x) + g(x) u} + w(x) \leq 0.
\end{equation} 
A control $u$ is admissible, if it fulfills \eqref{eqn:decay-cond}, where the existence of an input $u$ that fulfills \eqref{eqn:decay-cond} is ensured by Assumption \ref{asm:decay-condition}.
However, it this input is not unique.
Inequality \eqref{eqn:decay-cond} can be rewritten as
\begin{equation} \label{eqn:dfn-betai}
	\begin{split}
	&\scal{\nabla_x V(x), f(x) + g(x) u} + w(x) \\
	&= \underbrace{\scal{\nabla_x V(x), f(x)} + w(x)}_{=: \beta_0(x; w)} + \sum_{i = 1}^m \underbrace{\scal{\nabla_x V(x), g_i(x)}}_{=: \beta_i(x)} u_i \\
	\end{split}
\end{equation}
to show how $\eps$ influences the set of admissible controls $\hat \U(x; w) := \{ u \in \U: \varphi(u,x; w) \leq 0 \}$, where $g_i$ are the columns of $g(x)$.
As a short-hand notation, $\beta_0(x)$ is written instead of $\beta_0(x;w)$ whenever it is clear which $w$ is considered.

Since $x$ is unknown, $\ball_\eps(x)$ is replaced by $\ball_{2 \eps}(\hat x)$ in the following considerations since $\ball_\eps(x) \subset \ball_{2 \eps}(\hat x)$ holds right after a measurement.
Thus, inequality \eqref{eqn:decay-cond} has to be considered for all $\tilde x \in \ball_{2 \eps}(\hat x)$, such that $\beta_i(\tilde x)$ has to be computed for all of these points.
However, due to the properties of $f$, $g$, $V$ and $w$ (cf. \eqref{eqn:system} and Asm. \ref{asm:decay-condition}), each $\beta_i$ as a composition of continuously differentiable functions is Lipschitz continuous on a compact set of $\R^n$.
This ensures that $\beta_i(\hat x)$ can be used to compute a bound on $\beta_i(x)$ based on the Lipschitz constant. 
For this sake, the corresponding compact set $\set V$ is determined, on which the Lipschitz constants $L_i$ are computed.
Consider the starting ball $\hat R := \norm{\hat x_0} + 2 \eps$ along with the maximum CLF value given as $\hat V := \max_{\norm{x} \leq \hat R} V(x)$.
Since a control is computed that ensures a decay of all points within $\ball_{2 \eps}(\hat x)$, the highest Lyapunov function value is obtained for $\tilde x \in \hat R$.
Let $\alpha_1, \alpha_2 \in \mathcal K_\infty$ be given such that $\alpha_1(\norm x) \leq V(x) \leq \alpha_2(\norm x)$ holds \cite{Khalil2002-nonlin-sys}.
An overshoot bound is given as $\hat R^\star := \alpha_1^{-1}(\hat V)$, which means that all states are located in this ball, if a stabilizing control action is used.
Thus, $\set V := \ball_{\hat R^\star}$ is defined and all states are located in $\set V$ for $t \geq 0$.
Furthermore, the existence of the solution can be concluded due to Peano's theorem.
Since $f$ and $g$ are continuously differentiable and thus also countinuous, and the fact that the states are bounded by $\mathbb V$ and the inputs are bounded by input constraints, it means that the right-hand side of \eqref{eqn:system} is bounded.

After the compact set $\set V$ is defined, the Lipschitz constants can be computed either analytically or numerically such that
\begin{equation} \label{eqn:Lipschitz-betai}
	\abs{\beta_i(x) - \beta_i(\hat x)} \leq L_i \norm{x - \hat x}
\end{equation}
holds for all $x, \hat x \in \set V$.
This equation is used to bound the unknown terms $\beta_i(x)$, which is done in the following lemma.
\begin{lem} \label{lem:bounds-for-beta-Lip}
	Consider $\beta_i(x)$ from \eqref{eqn:dfn-betai}, $\set V := \ball_{\hat R^\star}$ with $\hat R^\star$ as overshoot bound, and Lipschitz constants $L_i$ from \eqref{eqn:Lipschitz-betai} for $i \in \{ 0:m \}$.	
	Then, 
	\begin{equation}
		\beta_i(\hat x) - L_i \norm{x - \hat x} \leq \beta_i(x) \leq \beta_i(\hat x) + L_i \norm{x - \hat x}
	\end{equation}
	holds for all $x, \hat x \in \set V$.
\end{lem}
\begin{proof}
	Set $\eps:= \norm{x - \hat x}$.
	The Lipschitz condition reads as $\abs{\beta_i(x) - \beta_i(\hat x)} \leq L_i \norm{x - \hat x} \leq L_i \eps$.
	Therefore, the following bounds hold for $i \in \{ 0:m \}$:
	\begin{equation} \label{eqn:beta_min_beta_max_bounds}
		\begin{split}
			&\abs{\beta_i(x) - \beta_i(\hat x)} \leq L_i \eps \\
			&\Leftrightarrow - L_i \eps \leq \beta_i(x) - \beta_i(\hat x) \leq L_i \eps \\
			&\Leftrightarrow \beta_i(\hat x) - L_i \eps \leq \beta_i(x) \leq \beta_i(\hat x) + L_i \eps.
		\end{split}
	\end{equation}
\end{proof}
This means that it is enough to consider $\beta_i(\hat x)$ along with $\eps$ and $L_i$ instead of the unknown $\beta_i(x)$.
Therefore, instead \eqref{eqn:decay-cond} with unknown $x$, $2^{m + 1}$ inequalities in $\hat x$ are considered.

\begin{figure}
	\centering
	\includegraphics[width = 0.23\textwidth]{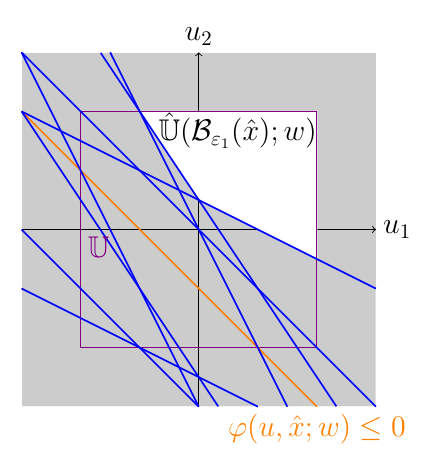}
	\includegraphics[width = 0.23\textwidth]{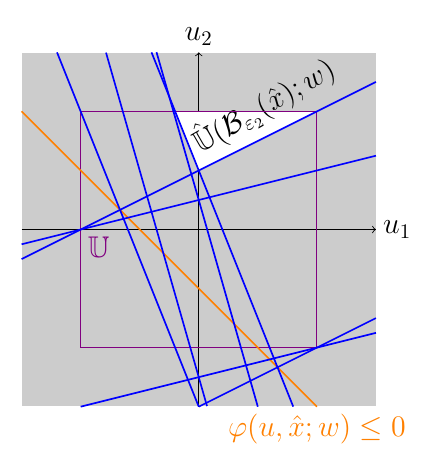}
	\caption{Visualization of $\varphi(u,\widehat x;w) \leq 0$ (orange) and the $2^{2+1} = 8$ inequalities (blue) that are obtained with the bounds in \eqref{eqn:beta_min_beta_max_bounds} as well as input constraints (violet). A control $u$ inside the white area is admissible, i.e. it ensures that $\varphi(u,x;w) \leq 0$ holds for all $x \in \ball_{\eps_i}(\widehat x)$. This area shrinks with growing $\eps$, where $\eps_1$ of the left figure is smaller than $\eps_2$ of the right figure.}
	\label{fig:hyperplanes-eps}
\end{figure}

Lemma \ref{lem:bounds-for-beta-Lip} provides a method to compute the unknown $\beta_i(x)$ and thus, it is possible to compute a state-dependent upper bound $\bar \eps(\hat x)$ for the required measurement accuracy such that the set of admissible controls $\hat{\U}(\ball_{\bar \eps(\hat x)}(\hat x); w) := \{ u \in \U: \varphi(u,x;w) \leq 0 \ \forall x \in \ball_{\bar \eps(\hat x)}(\hat x)\}$ is nonempty, i.e. $\forall \hat x \in \R^n \sm \{ 0 \} \ \exists \bar \eps(\hat x) > 0: \hat{\U} (\ball_{\bar \eps(\hat x)}(\hat x);w) \not= \emptyset$.
By satisfying $\varphi(u,x;w) \leq 0$ for all possible states, i.e. all points that can be reached by an arbitrary input (that satisfies the input constraints) starting in $\ball_{2 \eps}(\hat x)$, ensures stability for \eqref{eqn:system}.
The influence of the measurement accuracy on the set of admissible controls is shown for $m = 2$ in Fig. \ref{fig:hyperplanes-eps}: the higher $\eps_\cdot$ becomes, the smaller becomes the set of admissible controls $\hat{\U}(\ball_{\eps_\cdot}(\hat x);w)$, which is bounded by $\beta_i(x)$.
Additionally, it can be seen that $w$ only causes a parallel shift of $\varphi(u,\hat x;w)$, such that the set of admissible controls $\hat{\U}(\ball_{\eps_\cdot}(\hat x);w)$ becomes larger the smaller $w$ is.
This means that more controls are admissible as the demanded decay becomes smaller.

The aim of the following subsections is finding such a state-dependent bound $\bar \eps(x)$ to ensure a decay of all points around $\bar \eps(x)$ by the same input.
First, the approach is introduced and motivated.

\subsection{Visualization of required measurement accuracy}

Assumption \ref{asm:decay-condition} bounds all $\beta_i(\hat x)$ for a given $\hat x$.
Consider the case $m = 1$, i.e. $\beta_0(\hat x; w)$ and $\beta_1(\hat x)$ are computed for the given $\hat x$.
On the one hand, if $\beta_1(\hat x) = 0$ holds, then $\beta_0(\hat x; w) \leq 0$ holds, i.e. $\scal{\nabla_x V(\hat x), f(\hat x)} \leq -w(\hat x)$.
On the other hand, $\beta_0(\hat x; w) \geq 0 \implies \beta_1(\hat x) \not= 0$ holds. 
One problem is, that there might exist some $\hat x^\star \not= 0$ such that $\beta_0(\hat x^\star; w) = \beta_1(\hat x^\star) = 0$ holds, which means that even the smallest measurement error results in a violation of the decay condition.
Therefore, a relaxation is done, where some $\tilde w$ is considered for which $\varphi(u,x; \tilde w) \leq 0$ holds.
One suitable choice is $\tilde w = \alpha w$ for $\alpha \in (0,1)$, such that there is still a decay.
This relaxation ensures that there exists no $\hat x^\star \in \R^n \sm \{ 0 \}$ such that $\beta_0(\hat x^\star; \tilde w) = \beta_1(\hat x^\star) = 0$ holds.
It can be easily verified by assuming that such an $\hat x^\star$ exists.
If $\beta_0(\hat x^\star; \tilde w) = 0$, it means that $\scal{\nabla_x V(\hat x^\star), f(\hat x^\star)} = -\tilde w(\hat x^\star) \geq -w(\hat x^\star)$, which is a contradiction to \eqref{eqn:decay-condition-dot-V}.
Therefore, there exists no $x \in \R^n$ such that $\beta_0(x; \tilde w) \geq 0$ and $\beta_1(x) = 0$ hold simultaneously.
The decay $\tilde w$ serves as a margin from this ``prohibited'' axis in the sense that the smaller $\tilde w$ is at the current state, the larger is the distance to this axis. 
Thus, for all $x \in \R^n$, the points $(\beta_0(x; \tilde w), \beta_1(x))$ are located in an area that omits the positive part of the $\beta_0$ axis.
This fact is used for the computation of $\bar \eps$, where $\beta_0(x)$ is written instead of $\beta_0(x; \tilde w)$ for an ease of notation: 
Based on the measurement $\hat x$, $\beta_0(\hat x)$ and $\beta_1(\hat x)$ are determined.
First, consider the case that $\beta_0(\hat x) < 0$ and $\beta_1(\hat x) > 0$ hold. 
By enlarging the measurement error, there might exist $x_{0,-}$ and $x_{0,+}$ in $\ball_\eps(\hat x)$ with $\beta_0(x_{0,-}) < 0$ and $\beta_0(x_{0,+}) > 0$. 
Based on the derived Lipschitz constants $L_i$ (according to \eqref{eqn:Lipschitz-betai}), it means that $\beta_0(\hat x) - \eps L_0 \leq 0 \leq \beta_0(\hat x) + \eps L_0$ holds.
A control has to be determined to ensure $\dot V \leq 0$ especially for those $x$ for which $\beta_0(x) > 0$ holds, which is done via $\beta_1(x)$.
As long as $\beta_1(x)$ has the same sign for those points, an admissible $u$ can be chosen to ensure a decay.
On the other hand, by enlarging $\eps$, there might exist $x_{1,-}$ and $x_{1,+}$ such that $\beta_1(x_{1,-}) < 0$ and $\beta_1(x_{1,+}) > 0$ holds, which means that $\beta_1(\hat x) - \eps L_1 \leq 0 \leq \beta_1(\hat x) + \eps L_1$ holds. 
Since $\beta_0(\hat x) \leq 0$ in this case, $u = 0$ is admissible and $\eps$ can be enlarged as long as $\beta_0(x) \leq 0$ holds for all these points.
In short, $\eps$ can be enlarged as long as only the sign of one $\beta_i(x)$ is changing, where the input constraints $\set U$ have to be considered.
For $\beta_0(\hat x) > 0$ and $\beta_1(\hat x) > 0$, $\eps$ can be enlarged as long as there exists a point $x_{1,-}$ for which $\beta_1(x_{1,-}) < 0$ holds.
The other cases, i.e. for $\beta_1(\hat x) < 0$ are obtained simultaneously.

\begin{figure}
	\centering
	\includegraphics[trim={0.5cm 1.5cm 0.5cm 0},clip, width = 0.23\textwidth]{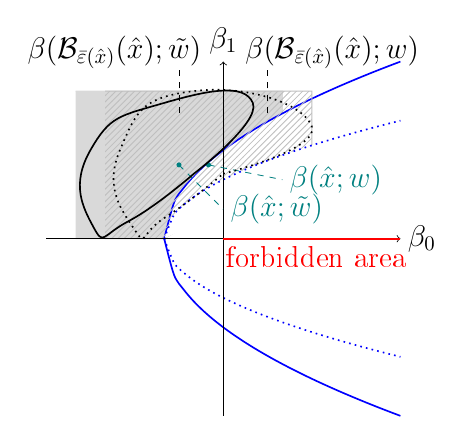}
	\includegraphics[trim={0.5cm 1.5cm 0.5cm 0},clip, width = 0.23\textwidth]{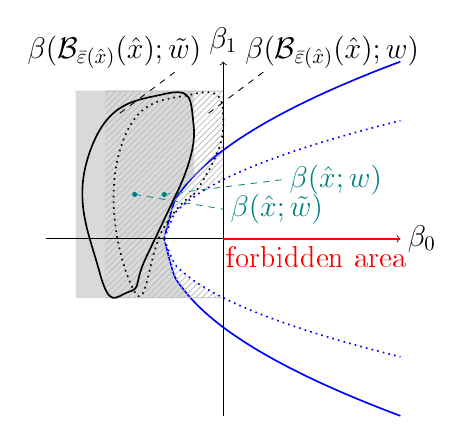}
	\includegraphics[trim={0.5cm 0 0.5cm 0},clip, width = 0.23\textwidth]{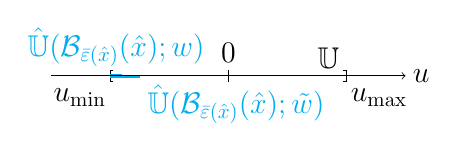}
	\includegraphics[trim={0.5cm 0 0.5cm 0},clip, width = 0.23\textwidth]{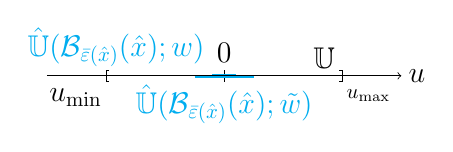}
	\caption{The effect of relaxing the decay condition. 
	Upper pictures:  For a required decay $w$ (blue dashed line) and a measured state $\widehat x$, $\beta_0(\widehat x; w)$ and $\beta_1(\widehat x)$ are computed. The values $\beta_0(x; w)$ and $\beta_1(x)$ of the unknown state $x$ are located in the light gray shaded area, which is given via Lemma \ref{lem:bounds-for-beta-Lip}.
	The maximum measurement error at $\hat x$ is determined such that this gray shaded area remains in maximum two quadrants.
	If the decay is relaxed to $\widetilde w$, $\beta_0(x; \widetilde w)$ shifts to the left along with the set where $\beta_0(x; \widetilde w)$ and $\beta_1(x)$ are located. 
	This relaxation of the decay might ensure that the maximum measurement error at $\widehat x$ can be enlarged.
	This case is presented on the right-hand side, where the shift to the left (caused by the relaxation of $w$) yields a buffer to the $\beta_1$ axis.
	It allows to enlarge the measurement error until the enlarged set touches the $\beta_1$ axis.
	Lower pictures: Due to the relaxation, the set of admissible controls enlarges.}
	\label{fig:relaxation}
\end{figure}

Based on this idea, the necessary measurement error $\bar \eps$ at each point $x \in \set V$ is computed.
Since $\eps$ is given, there might exist points where a measurement with a higher accuracy is required.
One of those domains is a vicinity around the origin, which depends on the decay rate $\tilde w$.
By ensuring $\beta_0(x) + \beta_1(x) u \leq 0$ with one $u \in \U$ for all possible states in a vicinity of the measurement $\hat x$, the set of admissible controls increases.
The effect of this relaxation is shown in Fig. \ref{fig:relaxation}, where $\beta(\ball_{\eps}(\hat x); w) := \{ \begin{pmatrix} \beta_0 (x; w) & \beta_1(x) \end{pmatrix}^\top: x \in \ball_\eps(\hat x) \}$.
It can be also seen that due to the relaxation of the decay condition, the measurement error can be again increased, which is also described in Fig. \ref{fig:relaxation}.
The remaining question is how far the measurement error can be increased. 
Therefore, the following subsection considers the connection between measurement error and decay condition.

\subsection{Computation of required measurement accuracy}

The presented thoughts are extended to $m > 1$ in the following theorem.
\begin{thm}
\label{thm:choice-of-eps}
	Consider system \eqref{eqn:system} with initial condition $x(0) = x_0$ together with input box constraints $\U$, where $0 \in \U$.
	Let Assumption \ref{asm:decay-condition} hold and let $\tilde w: \R^n \ra \R_{\geq 0}$ be given as the relaxed decay rate, which is continuously differentiable and satisfies $\tilde w(x) \leq w(x)$ for all $x \in \R^n$, $\tilde w(0) = 0$ and $x \not = 0 \implies \tilde w(x) > 0$.
	Additionally, let the initial measurement satisfy $\norm {\hat x_0 - x_0} \leq \eps$. 
	Then, 
	\begin{equation} \label{eqn:robustly-stabilizing}
		\begin{split}
			\forall \hat x \in \set V\sm \{ 0 \} \ &\exists \bar \eps(\hat x) > 0, \hat u \in \U: \\
			&\varphi(\hat u, x; \tilde w) \leq 0 \ \forall x \in \ball_{\bar \eps(\hat x)} (\hat x)
		\end{split}
	\end{equation}
	holds.
\end{thm}
	\begin{proof}
	The proof is divided into five parts.
	At first, a bound $\bar \eps_0(\hat x)$ is computed which ensures that $\beta_0(x) < 0$ holds for all $x \in \ball_{\bar \eps_0 (\hat x)}(\hat x)$ if $\beta_0(\hat x) \leq 0$.
	Second, a bound $\bar \eps_1(\hat x)$ is computed that ensures $\text{sign}(\beta_1(\hat x)) = \text{sign}(\beta_1(x))$ for all $x \in \ball_{\bar \eps_1 (\hat x)}(\hat x)$.
	The third part of the proof shows how $\bar \eps_0(\hat x)$ and $\bar \eps_1(\hat x)$ are used to compute $\bar \eps(\hat x)$ for a single input.
	In the fourth part, the results from the previous parts are extended to two inputs.
	Finally, the fifth and last part extends the results to more than two inputs.
	
	For the first three parts, a control $u$ has to be determined that fulfills all the four inequalities
	\begin{equation} \label{eqn:inequ_beta}
			\beta_0(\hat x) \pm L_0 \bar \eps_\cdot(\hat x) + (\beta_1(\hat x) \pm L_1 \bar \eps_\cdot(\hat x)) u \leq 0,
	\end{equation}
	since then, $\beta_0(x) + \beta_1(x) u \leq 0$ is satisfied (cf. Lemma \ref{lem:bounds-for-beta-Lip}) for all $x \in \ball_{\eps_\cdot}(\hat x)$.
	The Lipschitz constants are computed such that \eqref{eqn:Lipschitz-betai} holds.
	The arguments $\hat x$ and $\tilde w$ after $\beta_i$ and $\bar \eps_i$ are omitted in the proof.
	It is explicitly written, if new functions are introduced or a different argument is considered.	
	
	\textbf{Part 1:} Deriving $\bar \eps_0$.
	
	\underline{Case 1.1:} $\beta_0 \geq 0$: 
	For $\beta_0 = 0$, $\beta_1 \kappa \leq -w \leq -\tilde w$ holds due to Assumption \ref{asm:decay-condition}. 
	This means that $\beta_1 \not= 0$.
	Also for $\beta_0 > 0$,  $\beta_1 \not= 0$ holds.
	Therefore, there exists no $\bar \eps_0$ in this case, but some $\bar \eps_1$, which is considered in the second part of the proof.
		
	\underline{Case 1.2:} $\beta_0 < 0$: 
	In this case, the measurement error is chosen such that $\beta_0(x) \leq \beta_0 + L_0 \bar \eps_0 \leq 0$ is ensured.
	The bound follows as:
	\begin{equation} \label{eqn:bound-eps-0}
		\begin{split}
			\beta_0 + L_0 \eps \stackrel{!}{\leq} 0 \Leftrightarrow \eps \leq \bar \eps_0 := - \frac{\beta_0}{L_0}.
		\end{split}
	\end{equation}
		
	Due to the choice of $\bar \eps_0$, $\beta_0(x) \leq 0$ holds for all $x \in \ball_{\bar \eps_0(\hat x)}(\hat x)$.
	
	\textbf{Part 2:} Deriving $\bar \eps_1$ for one input.
	
	Now, the influence of an input $u$ is considered to compute a bound $\bar \eps_1$ for the measurement error.
	It might enlarge $\bar \eps_0$, which was computed in Case 1.2 for $\beta_0 < 0$.
	If $\beta_0 \geq 0$ holds, $\bar \eps_1$ is the only bound that has to hold, which exists since $\beta_1 \not= 0$ in these cases.
	Depending on the value of $\beta_1$, three cases are possible.
	
	\underline{Case 2.1:} $\beta_1 > 0$:
	Since $\beta_1 > 0$, a bound for $\bar \eps_1$ is derived such that $\beta_1 - L_1 \bar \eps_1 \geq 0$ holds.
	Therefore, $E_1^+ := \frac{\beta_1}{L_1}$ is introduced.
	The subscript results from the fact that $\beta_1$ appears on the right-hand side of $E_1^+$.
	The superscript is based on the sign of $\beta_1$.	
	However, the determined $E_1^+$ might require a control that does not fulfill the input constraints.
	To ensure them, an additional $E_{01}^+$ is introduced as
	\begin{equation} \label{eqn:eps_1_for_beta_1_geq_0}
		\begin{split}
			&\beta_0 + L_0 \bar \eps + \underbrace{(\beta_1 - L_1 \bar \eps)}_{\geq 0} \underbrace{u_{\min}}_{\leq 0} \leq 0 \\
			&\Leftrightarrow \bar \eps \leq E_{01}^+ := - \frac{\beta_0 + \beta_1 u_{\min}}{L_0 - L_1 u_{\min}} = - \frac{\beta_0 + \beta_1 u_{\min}}{L_0 + L_1 \abs{u_{\min}}}. \\
		\end{split}
	\end{equation}
	The determined $E_{01}^+$ fulfills all equations in \eqref{eqn:inequ_beta} since $(\beta_1 + L_1 \eps) u_{\min} \leq (\beta_1 - L_1 \eps) u_{\min} \leq 0$ and $\beta_0 - L_0 \eps \leq \beta_0 + L_0 \eps$ hold for all $\eps \geq 0$.
	Choosing $\bar \eps_1 := \min \left\{ E_{0}^+, E_{01}^+ \right\}$ ensures that $\beta_1(x) \geq 0$ for all $x \in \ball_{\bar \eps_1}(\hat x)$ holds and that an admissible control can be found to ensure \eqref{eqn:inequ_beta}.

	\underline{Case 2.2:} $\beta_1 < 0$:
	Analogously to Case 2.1,  $\bar \eps_1 = \min \{ E_1^-, E_{01}^- \}$ with $E_1^- := - \frac{\beta_1}{L_1}$ and $E_{01}^- := - \frac{\beta_0 + \beta_1 u_{\max}}{L_0 + L_1 u_{\max}}$ is chosen. 
	
	\underline{Case 2.3:} $\beta_1 = 0$:
	In this case, $\beta_0 \leq 0$ holds since there exists a stabilizing control for all $x \in \R$ based on Assumption \ref{asm:decay-condition}.
	Therefore, $\bar \eps_0$ is the only constraint that has to hold, which exists according to Case 1.2.	
	No additional bound $\bar \eps_1$ is obtained in this case.
		
	Finally, the three cases can be combined to obtain
	\begin{equation}
		\bar \eps_1 := \begin{cases}
		 \min \left\{ \frac{\beta_1}{L_1}, - \frac{\beta_0 + \beta_1 u_{\min}}{L_0 + L_1 \abs{u_{\min}}} \right\}, & \text{if } \beta_1 > 0 \\
		 \min \left\{ -\frac{\beta_1}{L_1}, - \frac{\beta_0 + \beta_1 u_{\max}}{L_0 + L_1 u_{\max}} \right\}, & \text{if } \beta_1 < 0 \end{cases},
	\end{equation}
	or shortly $\bar \eps_1 := \min \left\{ E_1, E_{01} \right\}$ with
	\begin{equation}
		E_1 := \frac{\abs{\beta_1}}{L_1}, E_{01} := - \frac{\beta_0 + \beta_1 \cdot \begin{cases} u_{\min}, & \beta_1 > 0 \\ u_{\max}, & \beta_1 < 0 \end{cases}}{L_0 + L_1 \cdot \begin{cases} \abs{u_{\min}}, & \beta_1 > 0 \\ u_{\max}, & \beta_1 < 0 \end{cases}}.		
\end{equation}		
	Keep in mind that all $E_\cdot$ are functions that depend on $\hat x$ and $\tilde w$.

	\textbf{Part 3:} Determining $\eps$ for one input.
	
	Two bounds $\bar \eps_0$ and $\bar \eps_1$ were derived in Part 1 and Part 2.
	The first bound $\bar \eps_0$ holds for $\beta_0 < 0$.
	The second bound $\bar \eps_1$ holds for $\beta_1 \not= 0$.
	Since there exists no $\hat x \not= 0$ such that $\beta_0 \geq 0$ and $\beta_1 = 0$ hold simultaneously, there exists at least one bound, which is chosen as $\bar \eps$.
	If $\hat x$ is measured such that two bounds exist, it is shown that $ \bar \eps := \max \{ \bar \eps_0, \bar \eps_1 \}$ yields a suitable choice as a bound for $\bar \eps$.
	Therefore, we consider the following three cases:

	\underline{Case 3.1:} $\bar \eps_0 > \bar \eps_1$:
	In this case, $\bar \eps$ yields that there might exist $x_+$ and $x_-$ where $\beta_1(x_-) < 0$ and $\beta_1(x_+) > 0$.
	However, $\beta_0(x) \leq 0$ is ensured for all $x \in \ball_{\bar \eps}(\hat x)$ due to the choice of $\bar \eps = \bar \eps_0$.
	Furthermore, $u = 0$ is admissible.
	Therefore, $\varphi(0,x; \tilde w) \leq 0$ holds for all $x \in \ball_{\bar \eps}(\hat x)$. 
	
	\underline{Case 3.2:} $\bar \eps_0 < \bar \eps_1$:
	In this case, there exist $x_+$ and $x_-$ where $\beta_0(x_-) < 0$ and $\beta_0(x_+) > 0$.
	Due to the choice of $\bar \eps = \bar \eps_1$, $\text{sign}(\beta_1(x)) = \text{sign}(\beta_1(\hat x))$ holds for all $x \in \ball_{\bar \eps}(\hat x)$.
	Therefore, there exists a $u \in \U$ such that $\varphi(u,x; \tilde w) \leq 0$ holds for all $x \in \ball_{\bar \eps}(\hat x)$ due to the choice of $\bar \eps = \bar \eps_1$.
	
	\underline{Case 3.3:} $\bar \eps_0 = \bar \eps_1$:
	This case can be considered as a special case of the Cases 3.1 and 3.2.

	\textbf{Part 4:} Deriving $\bar \eps_1$ for two inputs.
	
	In comparison to parts 1, 2, and 3, the interplay of the two controls $u_1$ and $u_2$ has to be investigated.
	There are different possibilities.
	First, $u_1$ is enough to stabilize the system at the current state, for example since $\beta_2$ is nearly zero and so $u_2$ has nearly no impact.
	Second, $u_2$ is enough to stabilize. 
	Third, either $u_1$ or $u_2$ is enough.
	Fourth, both controls are necessary.
	Those cases are considered now along with a case analysis of $\beta_1$ and $\beta_2$.
	According to the explanations of Part 1, for $\beta_0 \geq 0$ there exists at least one $i \in \{ 1,2 \}$ such that $\beta_i \not= 0$ holds and $\bar \eps_1 > 0$ is obtained as a bound.
	For $\beta_0 \leq 0$, no bound $\bar \eps_1$ is computed if $\beta_1 = \beta_2 = 0$ (cf. Case 2.3).

	\underline{Case 4.1.1:} $\beta_1 > 0, \beta_2 > 0$ and $u_1$ is enough to stabilize:
	In this case, $\beta_0 + \beta_1 u_{1,\min} \leq 0$ holds, but $\beta_0 + \beta_2 u_{2,\min} > 0$.
	It is possible to choose $\bar \eps_1$ such that either $u_2 = 0$ or or $u_2 \in [u_{2,\min}, u_{2,\max}] \sm \{ 0 \}$.	
	For $u_2 = 0$, $E_1^+$ and $E_{01}^+$ are obtained (cf. Case 2.1).
	Both bounds have to hold, such that the minimum of them is chosen.
	Choosing $u_2 \not= 0$ for stabilization, there exist upper bounds $E_1^+ := \frac{\beta_1}{L_1}$, $E_2^+ := \frac{\beta_2}{L_2}$, as well as
	\begin{equation} \label{eqn:bound-m-2-case-1}
		\begin{split}
			&\beta_0 + L_0 \bar \eps + (\beta_1 - L_1 \bar \eps) u_{1, \min} + (\beta_2 - L_2 \bar \eps) u_{2, \min} \leq 0 \\
			&\Leftrightarrow \bar \eps \leq E_{012}^{++} := - \frac{\beta_0 + \beta_1 u_{1, \min} + \beta_2 u_{2, \min}}{L_0 + L_1 \abs{u_{1, \min}} + L_2 \abs{u_{2, \min}}},
		\end{split}
	\end{equation}
	which follows from the same explanations as in Case 2.2 of the proof along with the explanation that $E_{012}^{++} > 0$ holds.
	Again, all three bounds have to be fulfilled, such that their minimum is chosen.
	To obtain the desired $\bar \eps_1$, the largest value of the two scenarios is obtained, since stabilization is possible either with or without $u_2$, i.e.	
	\begin{equation} \label{eqn:bounds-eps1-case-411}
		\bar \eps_1 = \max \left\{ \min \left\{ E_1^+, E_{01}^+ \right\}, \min \left\{ E_1^+, E_2^+, E_{012}^{++} \right\} \right\}.
\end{equation}				
	
	\underline{Case 4.1.2:} $\beta_1 > 0, \beta_2 > 0$ and $u_2$ is enough to stabilize:
	This case is simply Case 4.1.1 with changed indices, such that 
	\begin{equation}
		\bar \eps_1 = \max \left\{ \min \left\{ E_2^+, E_{02}^+ \right\}, \min \left\{ E_1^+, E_2^+, E_{012}^{++} \right\} \right\}
	\end{equation}			
is obtained, where $E_{02}^+$ is defined accordingly to $E_{01}^+$	.
	
	\underline{Case 4.1.3:} $\beta_1 > 0, \beta_2 > 0$ and either $u_1$ or $u_2$ is enough to stabilize:
	With the same argumentation as in the Cases 4.1.1 and 4.1.2, $\eps_1$ is obtained as
	\begin{equation}
		\begin{split}
			\bar \eps_1 = \max \left\{ \min \left\{ E_1^+, E_{01}^+ \right\}, \min \left\{ E_2^+, E_{02}^+ \right\}, \right. \\
			\left. \min \left\{ E_1^+, E_2^+, E_{012}^{++} \right\} \right\}.
		\end{split}
	\end{equation}		
	
	\underline{Case 4.1.4:} $\beta_1 > 0, \beta_2 > 0$ and both inputs are necessary to stabilize:	
	The desired bound reads as
	\begin{equation}
		\bar \eps_1 = \min \left\{ E_1^+, E_2^+, E_{012}^{++} \right\}.
	\end{equation}		
	
	These subcases can be combined such that $\eps_1$ is given as the maximum of $\min \left\{ E_1^+, E_2^+, E_{012}^{++} \right\}$ and all $\min \left\{ E_i^+, E_{0i}^+ \right\}$ for those $i = 1,2$, where $\beta_0 + \beta_i u_{i,\min} \leq 0$ hold. 
	
	\underline{Case 4.2:} $\beta_1 > 0, \beta_2 < 0$:
	The subcases are obtained in the same way as in the previous subcases, where the respective terms are replaced by $E_2^-$ and $E_{012}^{+-}$.
	The last term is obtained as
	\begin{equation}
		\begin{split}
			&\beta_0 + L_0 \bar \eps + (\beta_1 - L_1 \bar \eps) u_{1, \min} + (\beta_2 + L_2 \bar \eps) u_{2, \max} \leq 0 \\
			&\Leftrightarrow \bar \eps \leq E_{012}^{+-} := - \frac{\beta_0 + \beta_1 u_{1, \min} + \beta_2 u_{2, \max}}{L_0 + L_1 \abs{u_{1, \min}} + L_2 u_{2, \max}}.
		\end{split}
	\end{equation}	

	\underline{Case 4.3:} $\beta_1 > 0, \beta_2 = 0$: 
	This case was already investigated in the Case 2.1, i.e. $\bar \eps_1 := \min \left\{ E_1^+, E_{01}^+ \right\}$.
	
	All other cases for $\beta_1 \leq 0$ are obtained simultaneously. 
	The final choice of $\bar \eps_1$ can be shortly written as 
	\begin{equation}
		\bar \eps_1 = \max \left\{ \bar E_{012}, \bar E_{0i} \right\}
	\end{equation}
	for those $i \in \set I_2$ with $\bar E_{012} := \min \left\{ E_1, E_2, E_{012} \right\}$, $\bar E_{0i} := \min \left\{ E_i, E_{0i} \right\}$, where the index set $\set I_2$ is defined as
	\begin{equation} \label{eqn:index-set}
		\begin{split}
		\set I_m := \left\{ i \in \{1:m\}: \begin{cases} \beta_0 + \beta_i u_{i,\min} \leq 0, & \text{if } \beta_i > 0 \\ \beta_0 + \beta_i u_{i,\max} \leq 0, & \text{if } \beta_i < 0 \end{cases} \right\}
		\end{split}
	\end{equation}
	and 
	\begin{equation} \label{eqn:definition-E012}
		E_{012} = - \frac{\beta_0 + \sum \limits_{i = 1}^2 \beta_i \cdot \begin{cases} u_{i,\min}, & \beta_i > 0 \\ u_{i,\max}, & \beta_i < 0 \end{cases}}{L_0 + \sum \limits_{i = 1}^2 L_i \cdot \begin{cases} \abs{u_{i,\min}}, & \beta_i > 0 \\ u_{i,\max}, & \beta_i < 0 \end{cases}}.
	\end{equation}
	Finally, $\bar \eps$ is chosen as $\bar \eps_1$, if $\beta_0 \geq 0$ holds, or as $\bar \eps_0$, if $\beta_1 = 0$, where $\bar \eps_0 = -\frac{\beta_0}{L_0}$ (cf. Part 1).
	Otherwise, $\bar \eps := \max \{ \bar \eps_0, \bar \eps_1 \}$.
	
	\textbf{Part 5:} Deriving $\bar \eps_1$ for more than two inputs.
	
	The results of the last section are generalized to  more than two inputs, i.e. $m > 2$.
	This task is done recursively.
	Starting with $m = 3$, the index set $\set I_3$ is determined.
	
	\underline{Case 5.1:} $\abs{\set I_3} = 0$:
	In this case, each $u_i$ is necessary for stabilization and 
	\begin{equation}
		\bar \eps_1 = \bar E_{0123},
	\end{equation}
	where the functions $\bar E_{0123}$ and $E_{0123}$ are defined analogously to \eqref{eqn:definition-E012}.
	
	\underline{Case 5.2:} $\abs{\set I_3} = 1$:
	There exists one control, say $u_i$, which is necessary for stabilization.
	The others are optional.
	It means that either all three controls are used, i.e. $\bar E_{0123}$ is considered, or $u_i$ along with an optional control, i.e. $\bar E_{0ij}$ and $\bar E_{0ik}$ are considered, or $u_i$ alone, i.e. $\bar E_{0i}$ is considered.
	Therefore, the following choice is obtained:
	\begin{equation} \label{eqn:proof-part5}
		\bar \eps_1 = \max \left\{ \bar E_{0123}, \bar E_{0ij}, \bar E_{0ik}, \bar E_{0i} \right\}
	\end{equation}	
	for $i \in \set I_3$ and $j,k \not\in \set I_3$.
	For an ease of notation, $E_{0ij} = E_{0ji}$ is assumed in \eqref{eqn:proof-part5} and from now on.
	
	\underline{Case 5.3:} $\abs{\set I_3} =2$:
	If $\abs{\set I_3} = 2$, there exist two controls $u_i$ and $u_j$, that are necessary for stabilization.
	The remaining one is optional.
	Thus, 
	\begin{equation}
		\bar \eps_1 = \max \left\{ \bar E_{0123}, \bar E_{0ij}, \bar E_{0ik}, \bar E_{0jk}, \bar E_{0i}, \bar E_{0j} \right\}
	\end{equation}	
	is obtained based on the same explanations as in Case 5.2.
	
	\underline{Case 5.4:} $\abs{\set I_3} = 3$:
	In the last case, stabilization can be done with each control by itself.
	However, additional controls might enlarge $\eps_1$, such that the following expression is obtained:
	\begin{equation}
		\bar \eps_1 = \max \left\{ \bar E_{0123}, \bar E_{0ij}, \bar E_{0ik}, \bar E_{0jk}, \bar E_{0i}, \bar E_{0j}, \bar E_{0k} \right\}.
	\end{equation}	
	
	The following steps for $m > 3$ are just sketched to avoid a too nested expression.
	The procedure is always the same:
	Compute the index set $\set I_m$ and determine $\bar \eps_1$ based on its cardinality.
	To do so, combine all possible combinations between the $p$ indices $i_1, \ldots, i_p \in \set I_m$ and the $q  = m - p$ remaining indices $i_{p + 1}, \ldots, i_m \not\in \set I_m$ for a set of $1, 2, \ldots, \abs{\set I_m}$ used controls.
	Finally, the admissible measurement accuracy at each point $\hat x \in \set V$ is given as $\bar \eps := \max\{ \bar \eps_0, \bar \eps_1 \}$ with $\bar \eps_0 = -\frac{\beta_0}{L_0}$ for $\beta_0 \leq 0$, or simply as $\bar \eps_1$, if $\beta_0 > 0$.
\end{proof}

Since the maximum required measurement error at each point is determined, it can be used to exclude the states where the necessary measurement error is larger than the given $\eps$.
This prohibited area is denoted as core ball $\ball_{r^\star}$ and it ensures a decay outside this ball as long as $\norm x > r^\star$ holds, i.e. $\hat{\U} (\ball_{\eps}(\hat x); \tilde w) \not= \emptyset$ for all $\hat x \in \set V \sm \ball_{r^\star}$.
Before some conditions between the target ball and the core ball are introduced, a lemma is presented that introduces a lower bound on $\bar \eps(x)$ for all points outside a vicinity around the origin.

\begin{lem} \label{lem:lower-bound-eps}
	Consider system \eqref{eqn:system} with initial condition $x(0) = x_0$ together with input box constraints $\U$, where $0 \in \U$.
	Let Assumption \ref{asm:decay-condition} hold and let $\tilde w: \R^n \ra \R_{\geq 0}$ be given as the relaxed decay rate, which is continuously differentiable and satisfies $\tilde w(x) \leq w(x)$ for all $x \in \R^n$, $\tilde w(0) = 0$ and $x \not = 0 \implies \tilde w(\hat x) > 0$.
	Additionally, let the initial measurement satisfy $\norm {\hat x_0 - x_0} \leq \eps$ and let $\bar \eps(x)$ be computed as in the proof of Theorem \ref{thm:choice-of-eps}.
	Then, there exists a non-zero lower bound of $\bar \eps(\hat x)$ for each $\hat x \in \set V \sm \ball_{r^\star}$.
\end{lem}

\begin{proof}
	In the first part, the case $m = 1$ is considered (cf. Parts 1, 2, and 3 in the proof of Theorem \ref{thm:choice-of-eps}).
	In the second part, a generalization is done for $m > 1$ (cf. Parts 4 and 5 in the proof of Theorem \ref{thm:choice-of-eps}).
	
	\textbf{Part 1:} Lower bounds for $\bar \eps$ for $m =1$.
	
	\underline{Case 1.1:} $\beta_0 \geq 0$ and $\beta_1 > 0$:
	Due to Assumption \ref{asm:decay-condition} and the definition of $\tilde w$, $\beta_0(\hat x; \tilde w) + \beta_1(\hat x) u_{\min} \leq \beta_0(\hat x; \tilde w) + \beta_1(\hat x) \kappa \leq - w(\hat x) + \tilde w(\hat x)$ holds.
	Additionally, $- w(\hat x) + \tilde w(\hat x) \leq - \bar w$ can be ensured with
	 \begin{equation}
	 	\bar w := \min_{x \in \set V \sm \ball_{r^\star}} w(x) - \tilde w(x).
	\end{equation}	 	
	There exist lower bounds for $E_1^+$ and $E_{01}^+$ as $E_1^+ \geq \frac{\bar w}{L_1}$ and $E_{01}^+ \geq \frac{\bar w}{L_0 + L_1 \abs{u_{\min}}}$.
	Since $\bar \eps_1$ is chosen as the minimum of both bounds, its lower bound is given as $\frac{\bar w}{L_0 + L_1 \abs{u_{\min}}}$, since $\frac{\bar w + \beta_0}{L_1 \abs{u_{\min}}} \geq \frac{\bar w}{L_1 \abs{u_{\min}}} \geq \frac{\bar w}{L_0 + L_1 \abs{u_{\min}}}$ holds.
	
	\underline{Case 1.2:} $\beta_0 < 0$ and $\beta_1 > 0$:
	Beside the above derived bound for $\bar \eps_1$, there exists a bound $\bar \eps_0 = - \frac{\beta_0}{L_0}$.
	Since $\bar \eps$ is chosen as the maximum of $\bar \eps_0$ and $\bar \eps_1$, the lower bound can be simply chosen as $\frac{\bar w}{L_0 + L_1 \abs{u_{\min}}}$.
	
	\underline{Case 1.3:} $\beta_1 < 0$:
	Via the same considerations as in the two previous cases, $\frac{\bar w}{L_0 + L_1 \abs{u_{\max}}}$ is obtained as a lower bound for $\bar \eps$.
	
	\underline{Case 1.4:} $\beta_1 = 0$:
	Since $\beta_0 \not= 0$, $\frac{\bar w}{L_0}$ is obtained.
	
	To sum up the last cases, the lowest bound is given as $\frac{\bar w}{L_0 + L_1 u_{\text M}}$, where $u_{\text M} := \max\{ \abs{u_{\min}}, u_{\max} \}$.
	
	\textbf{Part 2:} Lower bounds for $\bar \eps$ for $m > 1$.
	
	Consider first the case $\beta_1 > 0$ and $\beta_2 > 0$.
	 As it was written in Case 4.1.1 in the proof of Theorem \ref{thm:choice-of-eps}, $u_1$ is enough to stabilize the system at the current $\hat x$, such that the bounds \eqref{eqn:bounds-eps1-case-411} hold. 
	 With the same explanations as in Part 1, the first minimum is bounded from below by $\frac{\bar w}{L_0 + L_1 \abs{u_{1,\min}}}$, whereas the second minimum is bounded from below by $\frac{\bar w}{L_0 + L_1 \abs{u_{1,\min}} + L_2 \abs{u_{2,\min}}}$. 
	 Since their maximum  is chosen, $\frac{\bar w}{L_0 + L_1 \abs{u_{1,\min}}}$ is a lower bound for $\bar \eps_1$ in this case.
	 
	 The remaining cases have a similar structure. 
	 To obtain a lower bound for the required measurement error without a lot of computations, the combinations of different controls is denied.
	 However, this yields a conservative choice, but a result that is easy to check, namely that the lower bound is given as
	 \begin{equation} \label{eqn:eps_min}
	 	\eps_{\min} := \frac 1 2 \frac{\bar w}{L_0 + \sum \limits_{i = 1}^m L_i u_{i, \text M}}, \ u_{i, \text M} := \max\{ \abs{u_{i, \min}}, u_{i, \max} \}.
	 \end{equation}
\end{proof}
	 The half in the fraction in \eqref{eqn:eps_min} is chosen since $\eps \leq \frac 1 2 \bar \eps$ has to hold.
	 It ensures that the stabilizing control law is admissible after a new measurement, since $\hat x \in \ball_\eps(x)$ holds and thus $x \in \ball_{2 \eps}(\hat x)$.
	 
\begin{rem}
	Instead of computing $\eps_{\min}$, $\bar \eps$ can be determined at each point $x \in \set V \sm \ball_{r^\star}$.
	This yields a less conservative bound, which is, however, more difficult to obtain.
\end{rem}

Since the lower bound of $\bar \eps$ outside the core ball $\ball_{r^\star}$ is known, it can be used as a condition for the measurement error.
This condition is presented in the next section along with the inter-execution time, i.e. the time between two measurements.

\section{Computation of next measurement time point} \label{sec:triggering-condition}

The previous section provided a required measurement accuracy at each point of the compact set $\set V$. 
By choosing $\eps$ as measurement error bound, there exist states where the measurement error is given as $\bar \eps(x) > 2 \eps$, i.e. the local measurement error is larger than the given error.
The larger the difference between $\bar \eps(x)$ and $\eps$ is, the larger is the time until the next measurement is necessary.
To determine a triggering condition for the next measuring time point, the evolution of the system is considered.
The following notation is introduced: 
By $\hat x^-(t_k)$ the state before the measurement is denoted and by $\hat x^+(t_k)$ the state after the measurement is denoted.

First of all, the dynamics of the system is denoted as $\dot x = f(x) + g(x) u =: F(x, u)$.
Its maximum is defined as
\begin{equation} \label{eqn:bar-F}
	\bar F := \sup_{x \in \set V \atop u \in \U} \norm{ F(x,u) }.
\end{equation}
The maximum is used to bound the distance between the state at $t = t_k$ and at the next measurement time point $t = t_k + \delta_k$ as
\begin{equation} \label{eqn:how-x-evolves}
	\begin{split}
		&\norm{x(t_k + \delta_k) - x(t_k)} = \norm{\int_{t_k}^{t_k +  \delta_k} F(x(\tau), u(\tau)) \ \mathrm d \tau} \\
		&\leq \int_{t_k}^{t_k + \delta_k} \norm{F(x(\tau), u(\tau))} \ \mathrm d \tau \leq \int_{t_k}^{t_k + \delta_k} \bar F \ \mathrm d \tau = \bar F \delta_k.
	\end{split}
\end{equation}
The so-called inter-execution time $\delta_k$ depends on the current state and can be determined by means of
\begin{equation} \label{eqn:inequality-bar-eps}
	\begin{split}
		&\norm{\hat x^+(t_k + \delta_k) - \hat x^+(t_k)} \\
		&\leq \norm{\hat x^+(t_k + \delta_k) - x(t_k + \delta_k)} \\
		& \ \ \ + \norm{x(t_k + \delta_k) - x(t_k)} + \norm{x(t_k) - \hat x^+(t_k)} \\
		&\leq \eps + \bar F \delta_k + \eps  = \bar F \delta_k + 2 \eps \stackrel ! \leq \bar \eps(\hat x^+(t_k)).
	\end{split}	 
\end{equation}
Rearranging the terms \eqref{eqn:inequality-bar-eps} yields the desired result such that $\delta_k$ is chosen as 
\begin{equation} \label{eqn:measuring-instants}
	\delta_k = \frac{\bar \eps(\hat x^+(t_k)) - 2 \eps}{\bar F}.
\end{equation}

\begin{figure}
	\centering
	\includegraphics[width = 0.5\textwidth, trim={0 0 0 3.3cm},clip]{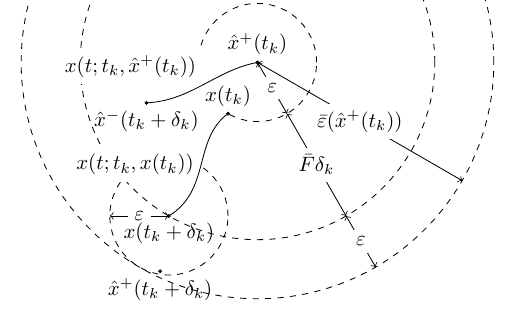}
	\caption{Visualization of \eqref{eqn:inequality-bar-eps}: Based on the current measurement, $\widehat x^+(t_k)$ is obtained. All points starting in $\ball_\eps(\widehat x^+(t_k))$ remain in $\ball_{\eps + \overline F \delta_k}(\widehat x^+(t_k))$. After the measurement after $\delta_k$ section, the measurement is located in $\ball_{2 \eps + \overline F \delta_k}(\widehat x^+(t_k))$. Since the maximum measurement error is computed as $\overline \eps(\widehat x^+(t_k))$, $\delta_k$ is chosen to satisfy this bound such that the set of admissible controls is nonempty after $\delta_k$ seconds and that the applied control is also admissible after a measurement.}
	\label{fig:feas-input}
\end{figure}

Based on the calculations, the set of admissible controls is empty for $t > t_k + \delta_k$.
At this point in time, a new measurement is necessary.
A graphical representation of the last inequality in \eqref{eqn:inequality-bar-eps} is shown in Fig. \ref{fig:feas-input}.

\begin{figure}
	\centering
	\includegraphics[width = 0.5\textwidth]{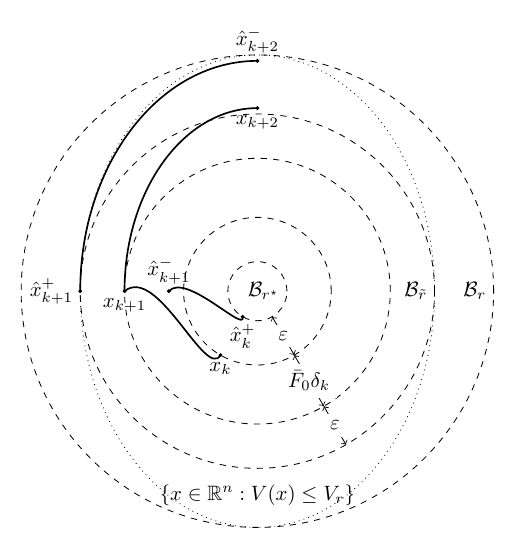}
	\caption{The procedure of the proof of Theorem \ref{thm:max-meas-err}: Once a measurement $\widehat x_k^+$ inside the core ball $\ball_{r^\star}$ is obtained, an arbitrary control, e.g. $u = 0$, is applied as long as it is ensured that the next measurement $\widehat x_{k+1}^+$ is located in the target ball $\ball_{\widetilde r}$. The set $\{ x \in \R^n: V(x) \leq V_r \}$ describes the level set of the CLF that the closed-loop trajectories do not leave since there exists a stabilizing control for all $\widehat x \in \ball_\eps(x_{k+1})$. Afterwards, the closed-loop trajectories evolve and remain in the target ball $\ball_r$.}
	\label{fig:core-ball-trajectories}
\end{figure}

Equation \eqref{eqn:measuring-instants} provides the measurement time points whenever the measured state is located outside the core ball.
To determine its size, consider the target ball $\ball_r$ and determine a triggering ball $\ball_{\tilde r}$, where a measurement is performed before the ball is left.
Its radius is given as 
\begin{equation} \label{eqn:core-ball-dfn}
	\tilde r := \alpha_2^{-1}(V_r), V_r = \alpha_1(r),
\end{equation}
where $V_r$ is given as the maximum value of the CLF that some $x \in \ball_{r}$ can obtain.
It holds since $\norm{x} \leq \alpha_2^{-1}(V_r) \implies V(x) \leq V_r$ and $V(x) \leq V_r = \alpha_1(r) \implies \norm x \leq r$ hold.
Thus, if the measured state is located in $\ball_{\tilde r}$, all closed-loop trajectories remain in $\ball_r$, since the exists a stabilizing control.
The desired core ball radius can be chosen to satisfy $r^\star + \eps < \tilde r - \eps$.
Since there exists not necessarily a control that stabilizes all possible states around the measurement that is located in the core ball, the distance between the core ball and the triggering ball yields an information how long an arbitrary control input can be applied until a new measurement is necessary.
To determine this time, $\bar F_0 := \sup \limits_{x \in \set V} \norm{F(x,0)}$ is determined as the maximum of the dynamics of the system by applying $u = 0$.
However, also a different control can be chosen for a measurement in the core ball, since $u$ can be chosen arbitrary in this ball.
Along with the triggering ball, the next measurement has to be executed at
\begin{equation} \label{triggering-in-target-ball}
	\delta_k := \frac{r - 2\eps - r^\star}{\bar F_0}.
\end{equation}	
Its choice is motivated in the following theorem, which presents the link between the required measurement accuracy, its lower bound, as well as the measurement time points.
\begin{thm}
\label{thm:max-meas-err}
	Consider system \eqref{eqn:system} with initial condition $x(0) = x_0$ together with input box constraints $\U$, where $0 \in \U$.
	Let Assumption \ref{asm:decay-condition} hold and let $\tilde w: \R^n \ra \R_{\geq 0}$ be given as the relaxed decay rate, which is continuously differentiable and satisfies $\tilde w(x) \leq w(x)$ for all $x \in \R^n$, $\tilde w(0) = 0$ and $x \not = 0 \implies \tilde w(x) > 0$.
	Let $\eps$ be given as the measurement error, let $\bar \eps(x)$ be defined as in the proof of Theorem \ref{thm:choice-of-eps} and let $\bar F$ be defined as \eqref{eqn:bar-F}.
	If $\eps < \eps_{\min}$ with 
	\begin{equation} \label{eqn:lower-bound-bar-eps}
		\begin{split}
			&\eps_{\min} = \frac 1 2 \frac{\bar w}{L_0 + \sum \limits_{i = 1}^m L_i u_{i, \text M}}, \ u_{i, \text M} := \max\{ \abs{u_{i, \min}}, u_{i, \max} \}, \\
			&\bar w := \min_{x \in \set V \sm \ball_{r^\star}} w(x) - \tilde w(x),
		\end{split}
	\end{equation}	
	holds, there exists a control law such that the following holds for $x(t; t_0, x_0)$, i.e. the closed-loop trajectory of \eqref{eqn:system}:
	For $\tilde r$ and $r$ satisfying \eqref{eqn:core-ball-dfn} as well as $r^\star$ satisfying $r^\star < \tilde r - 2 \eps$,	
	 the trajectory $x(t; t_0, x_0)$ enters the target ball $\ball_r$ and remains in it, if a new measurement is applied at
	\begin{equation} \label{eqn:triggering-time-points}
		\delta_k = \begin{cases} \frac{\bar \eps(\hat x^+(t_k)) - 2 \eps}{\bar F}, & \hat x^+(t_k) \not\in \ball_{r^\star} \\
		\frac{\tilde r - 2 \eps - r^\star}{\bar F_0}, & \hat x^+(t_k) \in \ball_{r^\star}
		\end{cases}.
\end{equation}		
\end{thm}
\begin{proof}
	The proof is divided into three parts.
	At first, the importance that \eqref{eqn:lower-bound-bar-eps} holds is motivated.
	Second, 	the convergence into $\ball_{r^\star}$ is shown.
	Third, it is shown that the closed-loop trajectory remains in $\ball_{r}$.
	
	\textbf{Part 1:}
	Lemma \ref{lem:lower-bound-eps} introduces the lower bound of $\eps$, which yields as a condition on the measurement accuracy of the sensor.
	Ensuring this condition lets $2 \eps \leq \bar \eps(x)$ hold for all $x \in \set V \sm \ball_{r^\star}$, i.e. for all $\hat x \in \set V \sm \ball_{r^\star}$ there exists a control $u$ such that $\varphi(u,x;\tilde w) \leq 0$ holds for all $x \in \ball_\eps(\hat x)$.
	
	\textbf{Part 2:}
	Theorem \ref{thm:choice-of-eps} provides a choice for $\bar \eps(\hat x)$.
	By ensuring it, the following property holds (cf. \eqref{eqn:robustly-stabilizing}) 
	\begin{equation} \label{eqn:decay-in-tk}
		\exists \hat u \in \U: \varphi(\hat u,x; \tilde w) \leq 0 \ \forall x \in \ball_{\bar \eps(\hat x^+_k)}(\hat x_k^+)
	\end{equation}	
	since $\bar \eps(x) > 2 \eps$ holds for all $x \in \set V \sm \ball_{r^\star}$.
		
	Let $\tilde x \in \ball_\eps(\hat x^+_k)$, i.e. $\norm{\hat x_k^+ - \tilde x} \leq \eps$.	
	Then according to \eqref{eqn:how-x-evolves} and the choice of $\delta_k$ as \eqref{eqn:measuring-instants} it holds that $\norm{\tilde x(t_k + \delta_k) - \tilde x(t_k)} \leq \bar F \delta_k \leq \bar \eps(\hat x^+_k) - 2\eps$.
	Therefore, $x(t; t_k, \tilde x_k) \in \ball_{\bar \eps(\hat x^+_k) - \eps}(\hat x^+_k)$ for all $t\in [0, \delta_k]$ and also $x_{k+1} := x(t_k + \delta_k) \in \ball_{\bar \eps(\hat x^+_k) - \eps}(\hat x^+_k)$.
	After a new measurement of $x_{k+1}$, $\norm{x_{k+1} - \hat x_{k+1}^+} \leq \eps$ holds and thus, $\hat x_{k+1}^+ \in \ball_{\bar \eps(\hat x^+_k)}(\hat x^+_k)$ such that \eqref{eqn:decay-in-tk} holds.
	At this point, the procedure starts anew until $\ball_{r^\star}$ is reached due to the minimum decay of $\tilde w$ at each point in time, since \eqref{eqn:decay-in-tk} holds for all states outside the core ball.
	
	\textbf{Part 3:}
	If $\hat x_k^+ \in \ball_{r^\star}$, there exists not necessarily an input such that $\varphi(u,x; \tilde w) \leq 0$ holds for all $x \in \ball_\eps(\hat x_k^+)$, i.e. no decay can be obtained for all those points.
	Since $\hat x_k^+ \in \ball_{r^\star} \implies x_k \in \ball_{r^\star + \eps}$, a control is determined such that $x_{k+1} \in \ball_{\tilde r - \eps}$ holds, which ensures $\hat x_{k+1}^+ \in \ball_{\tilde r}$.
	According to \eqref{eqn:how-x-evolves}, $\norm{x(t_k) - x(t_k + \delta_k)} \leq \bar F_0 \delta_k$ holds with $u = 0$.
	Since $\norm{x(t_k) - x(t_k + \delta_k)} \stackrel{!}{\leq} (\tilde r - \eps) - (r^\star + \eps) = \tilde r - 2\eps - r^\star$ has to be ensured such that $\hat x_{k+1}^+ \in \ball_{\tilde r}$ and $x_{k+1} \in \ball_{\tilde r - \eps}$, $\delta_k$	is chosen as
	\begin{equation}
		\bar F_0 \delta_k \leq \tilde r - 2\eps - r^\star \implies \delta_k := \frac{\tilde r - 2\eps - r^\star}{\bar F_0}.
	\end{equation}		
	These measurement time points ensure, that $x(t_k + \delta_k)$ remains in $\ball_{\tilde r - \eps}$. 
	If a new measurement is determined, $x^+(t_{k+1})$ is either still in $\ball_{r^\star}$, which is again Part 2, or $x^+(t_{k+1})$ is located in the triggering ball, but outside of the core ball, which is Part 1 and there exists a control such that all points in $\ball_\eps(\hat x^+_{k+1})$ can be stabilized.
	Due to the definition of the triggering ball according to \eqref{eqn:core-ball-dfn}, all closed-loop trajectories remain in the target ball since the maximum of the CLF is given as $V_r$.
	Since only these two cases can appear, $x(t; t_0, x_0)$ remains for all $t$ in the target ball after entering it once.
	The procedure is shown in Fig. \ref{fig:core-ball-trajectories}.
\end{proof}

\begin{rem}
	Due to the choice of the measurement accuracy, all trajectories starting in $\ball_{\eps}(\hat x_k)$ at $t = t_k$ never leave the current level set of the CLF
	\begin{equation} \label{eqn:dfn-V-k}
		\set V_k = \left\{ x \in \R^n: V(x) \leq \hat V_k:= \max_{\norm{\tilde x} \leq \hat R_k} V(\tilde x) \right\},
	\end{equation}		
	with $\hat R_k := \norm{\hat x_k} + 2 \eps$, such that the Lipschitz constants $L_0, L_1, \ldots, L_m$ can be computed again to obtain a larger $\bar \eps(\hat x^+_k)$ at each iteration, whenever $\hat x \not\in \ball_{r^\star}$.
\end{rem}

\begin{rem}
	Theorem \ref{thm:max-meas-err} provides a connection between the measurement error of the sensor $\eps$, the target ball $\ball_r$, the decay rate $w$ or $\tilde w$, and the inter-execution time $\delta_k$.
	As \eqref{eqn:lower-bound-bar-eps} and \eqref{eqn:triggering-time-points} suggest, decreasing $\tilde w$ results in a smaller decay and thus a more slowly convergence.
	However, the smaller $\tilde w$ is, the lower is the required measurement accuracy.
	Decreasing $\tilde w$ also comes along with less frequent measurements and smaller target and core ball, which means that the closed-loop trajectories converge closer to the origin.
	If the lower bound  \eqref{eqn:lower-bound-bar-eps} is not fulfilled, a sensor with a lower measurement error should be used.
	Otherwise, the radius of the target ball could be increased, which increases also the radius of the core ball along with the minimum decay $\bar w$.
	As a last possibility, the decay rate $\tilde w$ can be decreased, which results in a slower convergence to the target ball, but it increases also the inter-execution times $\delta_k$ such that measurements do not have to be taken to frequently. 
\end{rem}

Algorithm \ref{alg} summarizes the presented procedure.

\begin{algorithm}
	\caption{Stabilization under measurement errors}
	\label{alg}
	\begin{algorithmic}[1]
		\renewcommand{\algorithmicrequire}{\textbf{Input:}}
		\renewcommand{\algorithmicensure}{\textbf{Compute:}}
		\REQUIRE System $\dot x = f(x) + g(x) u$, Input constraints $\U$, CLF $V$, decay rate $w$, measurement accuracy $\eps$
		\STATE Measure the initial state $\hat x_0$
		\STATE Compute overshoot bound $\hat R^\star$
		\STATE Determine $\set V$ based on $\hat R^\star$
		\STATE Compute Lipschitz constants $L_0, L_1, \ldots, L_m$ on $\set V$ 
		\STATE Set target ball radius $r$ and compute radii of triggering ball $\tilde r$ and core ball $r^\star$
		\STATE Relax decay rate to $\tilde w$ and compute $\bar w$
		\STATE Compute the required measurement accuracy at each point, i.e. $\bar \eps(x)$ (cf. Thm. \ref{thm:choice-of-eps})
		\STATE Compute the lower bound of $\bar \eps(x)$ (cf. Lemma \ref{lem:lower-bound-eps}), i.e. $\eps_{\min}$
		\STATE Check if $\eps < \eps_{\min}$ holds
		\WHILE {system is running}
			\STATE Measure the state to obtain $\hat x_k$
			\STATE Update $\set V, L_0, L_1, \ldots, L_m$
			\IF {$\norm{\hat x_k} > r$}
				\STATE Compute $\delta_k$ (cf. \eqref{eqn:triggering-time-points})
				\STATE Compute the set of admissible controls $\hat{\U}(\ball_{\bar \eps(x)}; \tilde w)$
				\STATE Apply a feasible control for $\delta_k$ seconds
			\ELSE
				\STATE Compute $\delta_k$ (cf. \eqref{eqn:triggering-time-points})
				\STATE Apply arbitrary control, e.g. $u = 0$
			\ENDIF
		\ENDWHILE
	\end{algorithmic}
\end{algorithm}

The following section considers the application of Theorem \ref{thm:max-meas-err} on examples.
 
\section{Case study} \label{sec:casestudy}

Two examples are considered to show how the presented approach works.

\subsection{First example with one input}

A simplified train motion model is considered, which results from a balance of force equation as (cf. \cite{Hauck2023-train})
\begin{equation}
	\label{eq:model_train_2}
	\dot x = - \frac{F_\text{res}(x)}{m} + \frac{F_\text{train}(x)}{m} u,
\end{equation}
where $m$ is the mass of the train and $x$ is its velocity, which serves as the state of the system.
The resistance force is given as
\begin{equation}\label{eq:F_resistance}
	F_\text{res}(x) = p (x - v_\text{W})^2 + q
\end{equation}
and depends on air resistance (first term), as well as inclination resistance and rolling-resistance (second term), where $p, q > 0$ are assumed as constant.
The driving lever position is the input of the system and it is scaled to $[-1,1]$, by convention, where a positive value results in acceleration and a negative one in braking, i.e., $\U = [-1,1]$.
The train-specific driving power is given as
\begin{equation*}
	F_\text{train}(x) =  k_1 e^{- k_2 x} + k_3
\end{equation*}
with $k_i > 0$.
The aim is to accelerate the train from a measured speed $\hat x_0 = \SI{27}{\meter \per \second}$ to $x^\star = \SI{30}{\meter \per \second}$ at an inclination of one degree with wind speed $v_\text{W} = \SI{5}{\meter \per \second}$.
The parameters $p$ and $q$ are given based on the values in \cite{Hauck2023-train} as $p = \SI{5.18}{\kilogram \per \meter}$, $q = \SI{13046.32}{\newton}$, $m = \SI{68200}{\kilogram}$, $k_1 = \SI{1.516e+05}{\kilogram \meter \per \second^2}$, $k_2 = \SI{0.1147}{1 \per \second}$, $k_3 = \SI{1.564e+04}{\kilogram \meter \per \second^2}$.

The control law is designed to obtain $\dot x = 0$ at $x = x^\star$, such that $u^\star = \frac{ F_\text{res}(x^\star)}{F_\text{train}(x^\star)}$ has to be applied to hold the velocity at $x = x^\star$.
To ensure $\kappa \in \U$, it is chosen as $\kappa(x) = - \text{tanh}(x - \text{atanh}(u^\star) - x^\star)$.
Choosing $V = \frac 1 2 (x - x^\star)^2$ yields 
\begin{equation}
	\begin{split}
		\dot V = \scal{\nabla_x V, f + g \kappa} \leq - 0.025 (x - x^\star)^2 =: -w(x).
	\end{split}
\end{equation}
The last inequality was computed numerically based on the physical bounds of the considered train ($v_{\max} < \SI{40}{\meter \per \second}$).

The given sensor measurement accuracy is chosen as $\eps = 0.03$ and the relaxed decay is given as $\tilde w(x) = 0.015 (x - x^\star)^2$.
The target ball radius is chosen as $r = 1$ and the core ball radius as $r^\star = 0.9$, which means that a decay of the CLF can be ensured as long as the velocity is larger than $x^\star = \SI{30.9}{\meter \per \second}$ or less than $x^\star = \SI{29.1}{\meter \per \second}$.
A smaller $r^\star$ would result in more frequent measurements and a smaller required measurement accuracy.
Based on the given values, the required measurement accuracy $\eps_{\min}$ is computed as in \eqref{eqn:lower-bound-bar-eps} and yields $\eps_{\min} = 0.006$.
Thus, based on Theorem \ref{thm:max-meas-err}, practical stabilization can not be guaranteed.
However, the computed $\eps_{\min}$ is quite conservative. 
On the contrary, computing $\bar \eps$ at each point of the state space and choosing its minimum outside the core ball $\ball_{r^\star}$ as $\eps$, the required measurement accuracy is given as $0.04$. 
Since the sensor measurement error is smaller than this value, stabilization according to Theorem \ref{thm:max-meas-err} can be obtained.
The applied control is chosen as the midpoint of the set of admissible controls at the current measured state to ensure a compromise between  fast convergence to the desired value $x^\star$ and minimum acceleration to reach it.

\begin{figure}
	\centering
	\includegraphics[width = 0.47\textwidth]{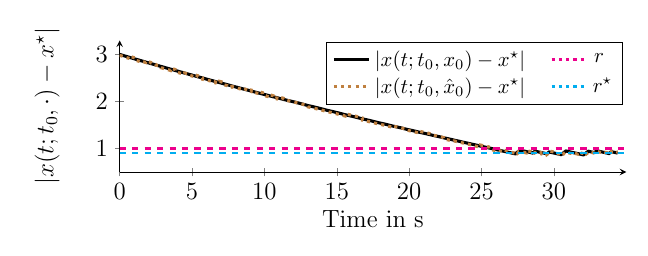}	
	\caption{Convergence of the closed-loop trajectories starting in $x_0$ and $\widehat x_0$ into the target ball $\ball_r$ with $r = 1$. 
	It can be seen that it remains in $\ball_{r}$ after entering it once. 
	The radius of the core ball is shown in cyan, while the radius of the target ball is added in magenta.}
	\label{fig:case_study_1_convergence}
\end{figure}

\begin{figure}
	\centering
	\includegraphics[trim={0cm 1cm 0cm 0cm},width = 0.47\textwidth]{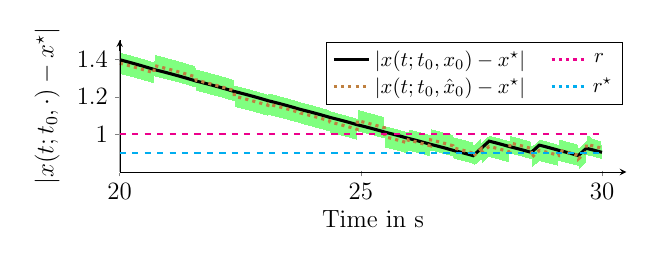}
	\includegraphics[width = 0.47\textwidth]{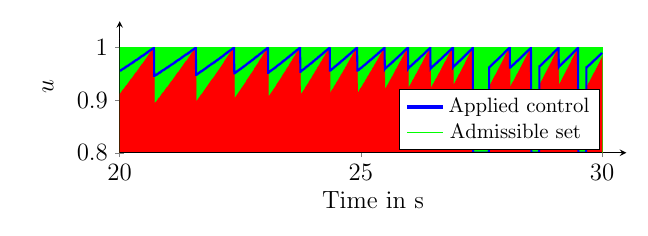}	
	\caption{Upper picture: Closer look of Fig. \ref{fig:case_study_1_convergence} for $t \in [20,30]$. 
	The green tube is the set where the closed-loop trajectories evolve after each measurement.	 
	Lower picture: Set of admissible controls (green) and the applied control (blue).		
	}
	\label{fig:case_study_1_results}
\end{figure}

Fig. \ref{fig:case_study_1_convergence} shows the convergence of the closed-loop trajectories into the target ball $\ball_r$.
It can be seen that after entering it once, the trajectories remain there.
A closer look of this convergence is shown in Fig. \ref{fig:case_study_1_results}.
The upper picture shows that all closed-loop trajectories that start in a vicinity of the measured state proceed in the green tube.
The presented approach ensures that all trajectories remain in the target ball, which can be seen since the green tube remains in $\ball_r$.
Whenever a measurement in the core ball $\ball_{r^\star}$ is obtained, $u = 0$ is applied as long as all closed-loop trajectories are located in the target ball.
Choosing $u = 0$ means that the train stops accelerating and thus, the velocity decreases and the states leave the core ball.
A new measurement is determined before the states leave $\ball_r$.
Due to the choice of the measurement time points, the measured state is still in $\ball_r$.
The applied control is shown in the lower picture.
It can be seen that the set of admissible controls shrinks the closer the trajectories reach the desired value $x^\star$, which means, that a higher driving lever position and a higher power has to be applied to reach this value.
It can be also seen that the inter-sampling time becomes smaller the closer the closed-loop trajectories approach the core ball $\ball_{r^\star}$.
A new measurement of the state is obtained before this set becomes zero.
The applied control is added in red and it is simply given as the mean of lower bound and upper bound of the set of feasible controls.
At the three time intervals at around 27.5, 28.5, and 29.5 seconds, $u = 0$ is chosen since the measurement is located in the core ball.

\subsection{Second example with two input}

As a second example, the following input-affine system is considered: 
\begin{equation} \label{eqn:system-case-study-2}
	\begin{split}
        \dot x = &\begin{pmatrix}
               - \frac{5}{4} x_2 - \frac{5}{10} x_3 - \frac{1}{16} \left(2 x_1 + x_2\right)^3 \\
               \frac{9}{10} x_1 + \frac{7}{10} x_2 + \frac{9}{10} x_3 \\
               - \frac{1}{2} x_1 - \frac{11}{8} x_2 - \frac{1}{4} x_3 - \frac{1}{32} \left(x_2 + 2 x_3\right)^3 
        \end{pmatrix}  \\
        &+ \begin{pmatrix}
               1 & 0 \\
               0 & 0 \\
               0 & 1
        \end{pmatrix} u.
        \end{split}
\end{equation}
Let $\U = [-1,1] \times [-0.5,0.5]$ be the set of input constraints.
It is aimed to practically stabilize \eqref{eqn:system-case-study-2} at the origin, which is a non-stable equilibrium point of \eqref{eqn:system-case-study-2}.
Using the quadratic control Lyapunov function
\begin{equation}
        V(x) = \frac 1 2 x^\top P x, P = \begin{pmatrix}
               1 & \frac 1 2 & 0 \\ \frac 1 2 & \frac 3 2 & \frac 1 2 \\ 0 & \frac 1 2 & 1
        \end{pmatrix} = \begin{pmatrix}
               p_1^\top \\ p_2^\top \\ p_3^\top
        \end{pmatrix},
\end{equation}
the control
\begin{equation}
        \kappa(x) = \begin{pmatrix}
               - \tanh(p_1^\top x) \\
               - \frac 1 2 \tanh(p_3^\top x)
        \end{pmatrix}
\end{equation}
asymptotically stabilizes \eqref{eqn:system-case-study-2}.
To verify this claim, the transformed coordinates $\tilde x_i := p_i^\top x$, $i =1,2,3$ are introduced.
The following calculation holds after rearranging the terms in \eqref{eqn:system-case-study-2}:
\begin{equation}       
        \begin{split}
               \dot V(x) &= \scal{\nabla_x V(x), f(x) + g(x) \kappa(x)} \\
               &= \tilde x_1 \left( \frac 1 2 \left( \tilde x_1 - \tilde x_1^3 \right) - \tilde x_2 - \tanh(\tilde x_1) \right) \\
	           & \ \ \ + \tilde x_2 \left( (\tilde x_1 + \tilde x_3) - \frac 1 5 \tilde x_2 \right) \\
	            & \ \ \ + \tilde x_3 \left( \frac 1 4 \left( \tilde x_3 - \tilde x_3^3\right) - \tilde x_2 - \frac 1 2 \tanh(\tilde x_3) \right) \\
	            &=  \frac 1 2 \left( \tilde x_1^2 - \tilde x_1^4 \right) - \tilde x_1 \tanh(\tilde x_1)
	              - \frac 1 5 \tilde x_2^2 \\
	              & \ \ \ + \frac 1 4 \left( \tilde x_3^2 - \tilde x_3^4 \right) - \frac 1 2 \tilde x_3 \tanh(\tilde x_3) \\
	            &\leq - \frac 1 4 \tilde x_1^2 - \frac{1}{10} \tilde x_2^2 - \frac 1 8 \tilde x_3^2 \\
				&= - \frac 1 2 \tilde x^\top \begin{pmatrix}
						\frac 1 2 & 0 & 0 \\
						0 & \frac{1}{5} & 0 \\
						0 & 0 & \frac 1 4
					\end{pmatrix} \tilde x \\
			    &= -\frac 1 2 x^\top \underbrace{P^\top \begin{pmatrix}
			    		\frac 1 2 & 0 & 0 \\
						0 & \frac{1}{5} & 0 \\
						0 & 0 & \frac 1 4
					\end{pmatrix} P}_{=: Q} x = - w(x). \\
        \end{split}
\end{equation}
The last inequality is obtained since $\frac 1 2 (y^2 - y^4) - y \tanh(y) \leq - \frac 1 2 y^2$ holds for $y \in \R$.
Instead of $w$, the relaxed decay $\tilde w = \frac 1 2 w$ is considered for the following computations.

The given sensor measurement accuracy is chosen as $\eps = 10^{-3}$.
The target ball radius is chosen as $r = 0.7$.
Based on \eqref{eqn:core-ball-dfn}, the triggering ball radius is computed as $r^\star = 0.347$ based on $\alpha_1 (r) = \lambda_{\min}(P) r^2$ and $\alpha_2 (r) = \lambda_{\max}(P) r^2$, where $\lambda_{\min}(P)$ is the minimum eigenvalue of $P$ and $\lambda_{\max}(P)$ is the maximum eigenvalue of $P$.
Defining the core ball radius as $r = 0.3$, yields the required measurement accuracy as $\eps_{\min} = 3.5 \cdot 10^{-4}$ as in \eqref{eqn:lower-bound-bar-eps}.
Since $\eps_{\min}$ is quite conservative, it is shown that practical stabilization can be also achieved by $\eps = 10^{-3}$.
Again, a less conservative value for $\eps_{\min}$ can be achieved by computing the required measurement accuracy at each point in the state space.
However, this method is computationally expensive for three dimensions.

\begin{figure}
        \centering
       \includegraphics[trim={0cm 1.2cm 0cm 0cm},width = 0.47\textwidth]{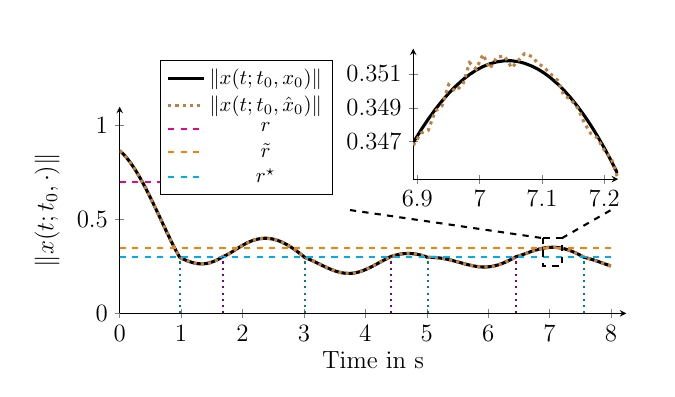}
       \includegraphics[width = 0.47\textwidth]{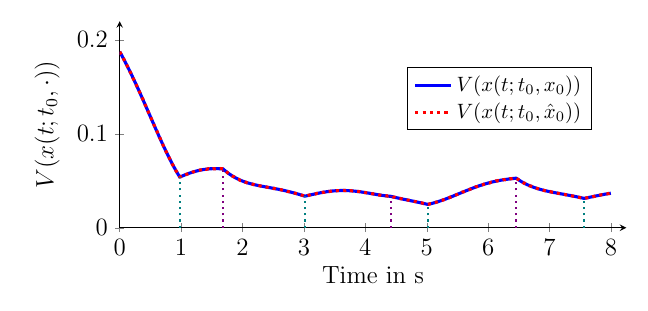}
        \caption{Upper picture: Norm of the closed-loop trajectories starting in $x_0$ and $\hat x_0$. Both trajectories remain in the target ball with radius $r = 0.7$ after entering it once. The picture in the upper right corner is a magnified view of the trajectories for $t \in [6.9, 7.2]$.
        Lower picture: Value of the CLF of the closed-loop trajectories.}
        \label{fig:case_study_2_results}
\end{figure}

The initial value is chosen as $x_0 = \frac 1 2 \begin{pmatrix} -1 & 1 & -1 \end{pmatrix}^\top$.
The upper picture in Fig. \ref{fig:case_study_2_results} shows the norm of the states, i.e., $\norm{x(t; t_0, x_0)}$ and $\norm{x(t; t_0, \hat x_0)}$, where $x(t; t_0, \hat x_0) := x(t; t_k, \hat x_k^+)$ for $t \in [t_k, t_{k+1})$.
The closed-loop trajectories converge into the target ball without an overshoot.
The applied control is chosen as the midpoint of the set of admissible controls at the current measured state.
However, since the algorithm computes the set of admissible controls, every  control in this set can be chosen instead.

The lower picture in Fig. \ref{fig:case_study_2_results} shows the value of the CLF evaluated at the closed-loop trajectories $x(t; t_0, x_0)$ and $x(t; t_0, \hat x_0)$.
It can be seen that the CLF value decreases until the trajectories reach the core ball at approx. $t_1 = 0.98$ (teal line).
After this time, there exists not necessarily an element in the set of admissible controls at the current measurement.
Therefore, $u = 0$ is chosen as long as the trajectories remain in the core ball, i.e. until approx. $t_2 = 1.68$ (violet line).
The lower picture shows also that the values of the CLF increase for $t \in [t_1,t_2]$.
For $t \in [t_2, t_3]$, with approx. $t_3 = 3.01$ (teal line) the value of the CLF decreases since a stabilizing control is applied, but the norm of the closed-loop trajectories increases, which is the result of the non-quadratic CLF.
However, the trajectories remain in the target ball $\ball_r$ due to the definition of $\tilde r$ and $r^\star$.
This procedure repeats for the remaining time and shows the effectiveness of the approach.

\subsection{Third example with two inputs}

The presented approach is now applied to the well-known Lotka-Volterra predator-prey model with controls
\begin{equation} \label{eqn:lotka-volterra}
	\dot x = \begin{pmatrix}
		\bar \alpha x_1 - \bar \beta x_2 x_1 \\
		- \bar \gamma x_2 + \bar \delta x_1 x_2
	\end{pmatrix} + \begin{pmatrix}
		x_1 & 0 \\
		0 & x_2
	\end{pmatrix} u.
\end{equation}
The Lotka-Volterra model describes interactions between species, namely a prey $x_1$ and a predator $x_2$.
The parameters were chosen as $\bar \alpha = 1.1, \bar \beta = 0.4, \bar \gamma = 0.4, \bar \delta = 0.1$ and describe the prey growth rate, prey death rate, predator death rate and predator growth rate.
The purpose of this example is stabilizing \eqref{eqn:lotka-volterra} in a vicinity of radius $r = 0.7$ at the fixpoint $x^\star = \begin{pmatrix} x_1^\star & x_2^\star \end{pmatrix}^\top = \begin{pmatrix} 10 & 4 \end{pmatrix}^\top$.
The procedure is considered for a set of initial values $x_0 \in \set X_0$, namely 
\begin{equation}
		\set X_0 := \left\{ \begin{pmatrix} 5 \\ 8 \end{pmatrix}, \begin{pmatrix} 10 \\ 6 \end{pmatrix}, \begin{pmatrix} 15 \\ 4 \end{pmatrix}, \begin{pmatrix} 10 \\ 3 \end{pmatrix}, \begin{pmatrix} 5 \\ 2 \end{pmatrix}, \begin{pmatrix} 1 \\ 3 \end{pmatrix}, \begin{pmatrix} 1 \\ 5 \end{pmatrix} \right\}
\end{equation}
The following function is a control Lyapunov function for \eqref{eqn:lotka-volterra} \cite{Meza2005-controller}: 
\begin{equation}
	V(x) =x_1 - x_1^\star - x_1^\star \log\left(\frac{x_1}{x_1^\star}\right) + x_2 - x_2^\star - x_2^\star \log\left(\frac{x_2}{x_2^\star}\right),
\end{equation}
which satisfies $V(x^\star) = 0$ and $V(x) > 0, \forall x \in \X \sm \{ x^\star \}$.
Together with this Lyapunov function, a feasible control according to Assumption \ref{asm:decay-condition} is given as
\begin{equation} \label{eqn:kappa-lotka-volterra}
		\kappa(x) = \begin{pmatrix}
			- \alpha + \beta x_2 + \tanh( - (x_1 - x_1^\star)) \\
			\gamma - \delta x_1 + \tanh( - (x_2 - x_2^\star))
		\end{pmatrix}.
\end{equation}
Using $\kappa(x)$, the derivative of $V$ results in
\begin{equation} \label{eqn:derivative-CLF-Lotka-Volterra}
	\begin{aligned}
		\dot V(x) =& -\frac 1 2  (x_1 - x_1^\star) \tanh( - (x_1 - x_1^\star)) \\
		&+ \frac 1 2 (x_2 - x_2^\star) \tanh( - (x_2 - x_2^\star)) =: -w(x),
	\end{aligned}
\end{equation}
which describes the decay rate of the controller $\kappa(x)$.
The relaxed decay rate is chosen as $\tilde w := \frac 1 2 w$
In the following, bounds for the input are determined. 
The bounds on the control can be derived from knowledge of the system.
Applying $u = 0$ on \eqref{eqn:lotka-volterra} results in a closed loop system, which can be bounded for the considered initial values $\set X_0$ by $[0,20] \times [0,8]$.
Due to these bounds, the optimal control \eqref{eqn:kappa-lotka-volterra} is bounded as $\kappa(x) \in [-3, 4] \times [-3, 2]$.

The given sensor measurement accuracy is chosen as $\eps = 10^{-2}$.
The target ball radius is chosen as $r = 0.7$.
Based on \eqref{eqn:core-ball-dfn}, the triggering ball radius is computed and set as $\tilde r = 0.3$ and the core ball radius as $r^\star = 0.2$.
These radii yield a required measurement accuracy as $\eps_{\min} = 1.3 \cdot 10^{-4}$ as in \eqref{eqn:lower-bound-bar-eps}.
Again, $\eps_{\min}$ is quite conservative and in the following, it is shown that even $\eps = 10^{-2}$ derives good results.
In comparison to the last two examples, the computation of the applied control is different. 
Again, the set of admissible controls is computed at each time instant.
Based on the determined $\hat{\set U}(x(t; t_k, \hat x_k); \tilde w)$, the control with the lowest costs is considered according to the objective function $J(u) = \frac 1 2 u^\top R u$ with $R = \text{diag}(3, 1)$, which means that influencing the prey is more costly then the predator.
Note that with $u \in \hat{\set U}(x(t; t_k, \hat x_k))$, a decay for all points around the measurement is ensured.

\begin{figure}   
        \centering
       \hspace*{1cm}
       \includegraphics[trim={2cm 1.6cm 0.5cm 0cm},clip,width = 0.35\textwidth]{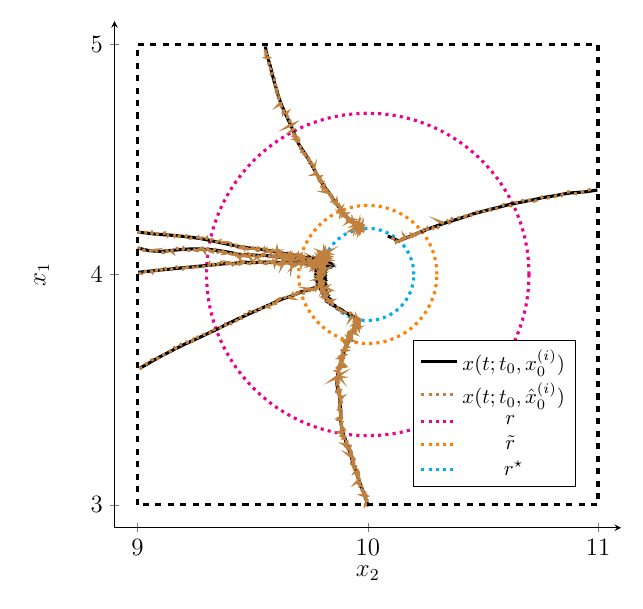}
       \includegraphics[trim={0cm 0cm 0cm 0.5cm},width = 0.47\textwidth]{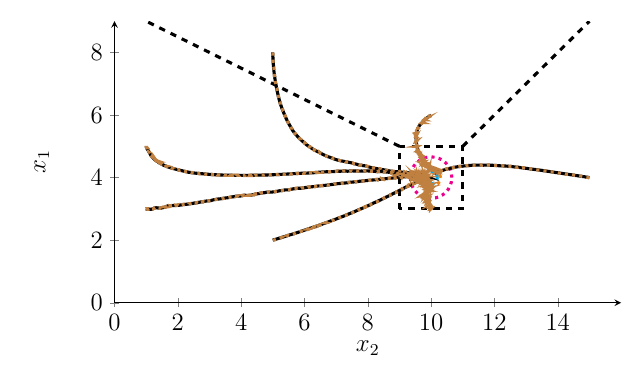}
        \caption{The closed-loop trajectories starting in the seven different initial values $x_o^{(i)} \in \set X_0$. All trajectories, i.e. the trajectories starting in the real values (black solid lines) and in the measured values (brown dotted lines), remain in the target ball with radius $r = 0.7$ after entering it once.
        The upper picture is a zoom of $[9, 11] \times [3, 5]$.}
        \label{fig:case_study_3_results}
\end{figure}

Fig. \ref{fig:case_study_3_results} shows the closed-loop trajectories starting at the seven different initial values $x_0^{(i)} \in \set X_0$.
All of them converge into the target ball $\ball_r(x^\star)$ and remain there.
Furthermore, some trajectories are moving along the boundary of $\ball_{r^\star}$, e.g. $x(t; t_0, \begin{pmatrix} 10 & 3 \end{pmatrix}^\top)$.
The reason is that $u = 0$ yields a movement of the trajectory to the left, such that it exits the core ball. 
Stabilizing it yields a movement upwards into the core ball. 
The other trajectories remain more or less at the same point, which results from the fact that the triggering time points that are obtained for states in the core ball $\ball_{r^\star}$ are conservative.
This even ensures that the trajectories remain in the triggering ball $\ball_{\tilde r}$.

\section{Conclusion} \label{sec:concl}

This paper considered stabilization in the presence of measurement uncertainties, where specific bounds for the measurement error are derived.
In the first part of this work, a bound for the maximum measurement error at each point of a compact set is derived, such that there exists an input that yields a decay of the CLF for all points in a vicinity of the measured state.
The decay was relaxed to increase this bound. 
Afterwards, the radius of the core ball was chosen, which is defined as the set, where a decay is ensured outside of this ball.
Based on this core ball and the relaxed decay rate, a lower bound of the maximum measurement error outside the core ball was computed.
A new measurement is performed whenever the discrepancy between the measured state and the states in the ball violate a certain threshold.
Additionally, a theorem that shows the convergence of the closed-loop trajectories into the target ball, in which they remain, was presented in the second part. 
Furthermore, admissibility of the control after a new measurement is ensured.
In two examples, the effectiveness of the approach was shown.

Future work includes the extension to general input constraints as well as considering lower bounds for the inter-execution time to ensure a minimum time between two sensor measurements.
Furthermore, investigations to ensure the existence of a continuously differentiable control as a result of the procedure provide further insight and increase the applicability of the approach.
Additionally, considering delays in either the transmission of the sensor data or the computation of the control are an interesting and relevant topic \cite{Cheddour2023-feedback}, which appears quite often in practice.
The presented work can be combined with approaches that require the set of admissible controls to determine a control which is optimal w.r.t. a given objective function, e.g. in MPC or time-varying optimal control \cite{Schmidt2021-tracking}.


\section*{Acknowledgment}

The authors thank Thomas G\"ohrt for his valuable comments on the publication over the last few months.
This research was funded by the German Ministry for Education and Research (BMBF) in the frame of the SRCC$\_$EETCM project, grant number 03WIR1208.
The publication of this article was funded by Chemnitz University of Technology.

\bibliographystyle{plain}
\bibliography{bib/triggered}

\begin{thebibliography}{10}

\bibitem{Anta2010-sample}
A.~Anta and P.~Tabuada.
\newblock To sample or not to sample: Self-triggered control for nonlinear
  systems.
\newblock {\em IEEE Transactions on automatic control}, 55(9):2030--2042, 2010.

\bibitem{Baier2012-linear}
R.~Baier, L.~Gr{\"u}ne, and S.F. Hafstein.
\newblock Linear programming based {L}yapunov function computation for
  differential inclusions.
\newblock {\em Discrete and Continuous Dynamical Systems-Series B},
  17(1):33--56, 2012.

\bibitem{Baier2014-num-CLF}
R.~Baier and S.F. Hafstein.
\newblock Numerical computation of control {L}yapunov functions in the sense of
  generalized gradients.
\newblock In {\em Proceedings of the 21st International Symposium on
  Mathematical Theory of Networks and Systems}, pages 1173--1180, 2014.

\bibitem{Behera2014-event}
A.~K. Behera and B.~Bandyopadhyay.
\newblock Event based robust stabilization of linear systems.
\newblock In {\em IECON 2014-40th Annual Conference of the IEEE Industrial
  Electronics Society}, pages 133--138. IEEE, 2014.

\bibitem{Bianchini2018-merging}
F.~Bianchini, F.~Fabiani, and S.~Grammatico.
\newblock On merging constraint and optimal control-{L}yapunov functions.
\newblock In {\em 2018 IEEE Conference on Decision and Control (CDC)}, pages
  2328--2333. IEEE, 2018.

\bibitem{Borgers2018-time}
D.P. Borgers, V.S. Dolk, G.E. Dullerud, A.R. Teel, and W.P.M.H. Heemels.
\newblock Time-regularized and periodic event-triggered control for linear
  systems.
\newblock {\em Control Subject to Computational and Communication Constraints:
  Current Challenges}, pages 121--149, 2018.

\bibitem{Brunner2019-event}
F.~D. Brunner, W.P.M.H. Heemels, and F.~Allg{\"o}wer.
\newblock Event-triggered and self-triggered control for linear systems based
  on reachable sets.
\newblock {\em Automatica}, 101:15--26, 2019.

\bibitem{Cheddour2023-feedback}
A.~Cheddour, A.~Elazzouzi, and M.~Ouzahra.
\newblock Feedback stabilization of semilinear system with distributed delay.
\newblock {\em IEEE Transactions on Automatic Control}, 69(1):129--144, 2023.

\bibitem{Delimpaltadakis2021-region}
G.~Delimpaltadakis and M.~Mazo.
\newblock Region-based self-triggered control for perturbed and uncertain
  nonlinear systems.
\newblock {\em IEEE Transactions on Control of Network Systems}, 8(2):757--768,
  2021.

\bibitem{DiBenedetto2013-digital}
M.D. Di~Benedetto, S.~Di~Gennaro, and A.~D'Innocenzo.
\newblock Digital self-triggered robust control of nonlinear systems.
\newblock {\em International Journal of Control}, 86(9):1664--1672, 2013.

\bibitem{Giesl2015-review}
P.~Giesl and S.F. Hafstein.
\newblock Review on computational methods for {L}yapunov functions.
\newblock {\em Discrete and Continuous Dynamical Systems-Series B},
  20(8):2291--2331, 2015.

\bibitem{Gommans2015-resource}
T.M.P. Gommans and W.P.M.H. Heemels.
\newblock Resource-aware mpc for constrained nonlinear systems: A
  self-triggered control approach.
\newblock {\em Systems \& Control Letters}, 79:59--67, 2015.

\bibitem{Hauck2023-train}
M.~Hauck, P.~Schmidt, A.~Kobelski, and S.~Streif.
\newblock A map-based model predictive control approach for train operation.
\newblock In {\em Proc. of the 21st European Control Conference}, 2023.

\bibitem{Heemels2012-introduction}
W.P.M.H. Heemels, K.H. Johansson, and P.~Tabuada.
\newblock An introduction to event-triggered and self-triggered control.
\newblock In {\em 51st {IEEE} conference on decision and control ({CDC})},
  pages 3270--3285. IEEE, 2012.

\bibitem{Heemels2021-event}
W.P.M.H. Heemels, K.H. Johansson, and P.~Tabuada.
\newblock Event-triggered and self-triggered control.
\newblock In {\em Encyclopedia of Systems and Control}, pages 724--730.
  Springer, 2021.

\bibitem{Hertneck2021-dynamic}
M.~Hertneck and F.~Allg{\"o}wer.
\newblock Dynamic self-triggered control for nonlinear systems based on hybrid
  lyapunov functions.
\newblock In {\em 2021 60th IEEE Conference on Decision and Control (CDC)},
  pages 533--539. IEEE, 2021.

\bibitem{Hertneck2022-dynamic}
M.~Hertneck and F.~Allg{\"o}wer.
\newblock Dynamic self-triggered control for nonlinear systems with delays.
\newblock {\em IFAC-PapersOnLine}, 55(13):312--317, 2022.

\bibitem{Hertneck2023-self}
M.~Hertneck and F.~Allg{\"o}wer.
\newblock Self-triggered output feedback control for nonlinear networked
  control systems based on hybrid lyapunov functions.
\newblock {\em IFAC-PapersOnLine}, 56(2):5255--5260, 2023.

\bibitem{Hertneck2024-robust}
M.~Hertneck and F.~Allg{\"o}wer.
\newblock Robust dynamic self-triggered control for nonlinear systems using
  hybrid lyapunov functions.
\newblock {\em Nonlinear Analysis: Hybrid Systems}, 53:101485, 2024.

\bibitem{Khalil2002-nonlin-sys}
H.~Khalil.
\newblock {\em Nonlinear {S}ystems}.
\newblock Prentice-Hall. 3nd edition, 2002.

\bibitem{Kishida2023-greedy}
M.~Kishida.
\newblock Greedy synthesis of event-and self-triggered controls with control
  lyapunov-barrier function.
\newblock In {\em 2023 62nd IEEE Conference on Decision and Control (CDC)},
  pages 4467--4473. IEEE, 2023.

\bibitem{Kogel2015-robust}
M.~K{\"o}gel and R.~Findeisen.
\newblock Robust output feedback predictive control with self-triggered
  measurements.
\newblock In {\em 2015 54th IEEE Conference on Decision and Control (CDC)},
  pages 5487--5493. IEEE, 2015.

\bibitem{Ledyaev1997-remark}
Y.~S. Ledyaev and E.~D. Sontag.
\newblock A remark on robust stabilization of general asymptotically
  controllable systems.
\newblock In {\em Proc. Conf. on Information Sciences and Systems (CISS 97),
  Johns Hopkins, Baltimore, MD}, pages 246--251, 1997.

\bibitem{Li2019-event}
Y.~Li, L.~Liu, C.~Hua, and G.~Feng.
\newblock Event-triggered/self-triggered leader-following control of stochastic
  nonlinear multiagent systems using high-gain method.
\newblock {\em IEEE Transactions on Cybernetics}, 51(6):2969--2978, 2019.

\bibitem{Li2001-switched}
Z.~G. Li, C.~Y. Wen, and Y.~C. Soh.
\newblock Switched controllers and their applications in bilinear systems.
\newblock {\em Automatica}, 37(3):477--481, 2001.

\bibitem{Liberzon2005-stabilization}
D.~Liberzon and J.~P. Hespanha.
\newblock Stabilization of nonlinear systems with limited information feedback.
\newblock {\em IEEE Transactions on Automatic Control}, 50(6):910--915, 2005.

\bibitem{lin2024artificial}
H.~Lin, J.~Dong, and J.~H. Park.
\newblock An artificial-delay-based looped functional for dynamic
  event-triggered fault-tolerant control of ts fuzzy multi-agent systems.
\newblock {\em IEEE Transactions on Automation Science and Engineering}, 2024.

\bibitem{Liu2023-data}
W.~Liu, J.~Sun, G.~Wang, F.~Bullo, and J.~Chen.
\newblock Data-driven self-triggered control via trajectory prediction.
\newblock {\em IEEE Transactions on Automatic Control}, 68(11):6951--6958,
  2023.

\bibitem{Mazenc2019-stabilization}
F.~Mazenc, L.~Burlion, and M.~Malisoff.
\newblock Stabilization and robustness analysis for a chain of saturating
  integrators with imprecise measurements.
\newblock {\em IEEE control systems letters}, 3(2):428--433, 2019.

\bibitem{Mazo2009-self}
M.~Mazo, A.~Anta, and P.~Tabuada.
\newblock On self-triggered control for linear systems: Guarantees and
  complexity.
\newblock In {\em 2009 European Control Conference (ECC)}, pages 3767--3772.
  IEEE, 2009.

\bibitem{Meza2005-controller}
M.~E.~M. Meza, A.~Bhaya, and E.~Kaszkurewicz.
\newblock Controller design techniques for the lotka-volterra nonlinear system.
\newblock {\em Sba: Controle \& Automa{\c{c}}{\~a}o Sociedade Brasileira de
  Automatica}, 16:124--135, 2005.

\bibitem{Proskurnikov2019-lyapunov}
A.V. Proskurnikov and M.~Mazo.
\newblock Lyapunov event-triggered stabilization with a known convergence rate.
\newblock {\em IEEE Transactions on Automatic Control}, 65(2):507--521, 2019.

\bibitem{Rajan2023-analysis}
A.~Rajan and P.~Tallapragada.
\newblock Analysis of inter-event times in linear systems under region-based
  self-triggered control.
\newblock {\em IEEE Transactions on Automatic Control}, 2023.

\bibitem{Sabouni2024-optimal}
E.~Sabouni, C.G. Cassandras, W.~Xiao, and N.~Meskin.
\newblock Optimal control of connected automated vehicles with
  event/self-triggered control barrier functions.
\newblock {\em Automatica}, 162:111530, 2024.

\bibitem{Scheres2024-robustifying}
K.J.A. Scheres, R.~Postoyan, and W.P.M.H. Heemels.
\newblock Robustifying event-triggered control to measurement noise.
\newblock {\em Automatica}, 159:111305, 2024.

\bibitem{Schmidt2021-tracking}
P.~Schmidt, T.~G{\"o}hrt, and S.~Streif.
\newblock Tracking of stabilizing, optimal control in fixed-time based on
  time-varying objective function.
\newblock pages 6012--6017, 2021.

\bibitem{Schmidt2021-inf}
P.~Schmidt, P.~Osinenko, and S.~Streif.
\newblock On inf-convolution-based robust practical stabilization under
  computational uncertainty.
\newblock {\em IEEE Transactions on Automatic Control}, 66(11):5530--5537,
  2021.

\bibitem{Sontag1999-stabilization-disturb}
E.~Sontag.
\newblock Stability and stabilization: discontinuities and the effect of
  disturbances.
\newblock In {\em Nonlinear {A}nalysis, {D}ifferential {E}quations and
  {C}ontrol}, pages 551--598. Springer, 1999.

\bibitem{Tarbouriech2018-insights}
S.~Tarbouriech, A.~Seuret, C.~Prieur, and L.~Zaccarian.
\newblock Insights on event-triggered control for linear systems subject to
  norm-bounded uncertainty.
\newblock {\em Control Subject to Computational and Communication Constraints:
  Current Challenges}, pages 181--196, 2018.

\bibitem{Yi2018-dynamic}
X.~Yi, K.~Liu, D.V. Dimarogonas, and K.H. Johansson.
\newblock Dynamic event-triggered and self-triggered control for multi-agent
  systems.
\newblock {\em IEEE Transactions on Automatic Control}, 64(8):3300--3307, 2018.

\bibitem{Zhang2022-finite}
K.~Zhang, B.~Zhou, W.~X. Zheng, and G.-R. Duan.
\newblock Finite-time stabilization of linear systems by bounded
  event-triggered and self-triggered control.
\newblock {\em Information Sciences}, 597:166--181, 2022.

\bibitem{Zhang2021-new}
P.~Zhang, T.~Liu, and Z.-P. Jiang.
\newblock New results in stabilization of uncertain nonholonomic systems: A
  self-triggered control approach.
\newblock In {\em 40th Chinese Control Conference (CCC)}, pages 686--693. IEEE,
  2021.

\bibitem{Zhang2022-tracking}
P.~Zhang, T.~Liu, and Z.-P. Jiang.
\newblock Tracking control of unicycle mobile robots with event-triggered and
  self-triggered feedback.
\newblock {\em IEEE Transactions on Automatic Control}, 68(4):2261--2276, 2022.

\end{thebibliography}

\end{document}